\documentclass[sn-mathphys,Numbered]{sn-jnl}

\usepackage{bm}%
\usepackage{siunitx}%
\usepackage[version=4]{mhchem}%
\usepackage{graphicx}%
\usepackage{cancel}%
\usepackage{mathtools}%
\usepackage{enumitem}%
\usepackage{nameref}%
\usepackage{multirow}%
\usepackage{amsmath,amssymb,amsfonts}%
\usepackage{amsthm}%
\usepackage{mathrsfs}%
\usepackage[title]{appendix}%
\usepackage[dvipsnames]{xcolor}%
\usepackage{textcomp}%
\usepackage{manyfoot}%
\usepackage{booktabs}%
\usepackage{algorithm}%
\usepackage{algorithmicx}%
\usepackage{algpseudocode}%
\usepackage{listings}%
\usepackage{lineno}
\usepackage{soul}

\newcommand{\real}[1]{{\mathbb{R}^{#1}}}
\newcommand{\fancy}[1]{\mathscr{#1}}
\newcommand{\suppinfo}[1][]{see #1 in the Appendix}

\DeclarePairedDelimiterX{\infdivx}[2]{(}{)}{%
  #1\;\delimsize\|\;#2%
}

\newcommand{\expect}[2]{\mathbb{E}_{#1}[#2]}
\newcommand{\var}[2]{\mathbb{V}\mathrm{ar}_{#1}[#2]}

\theoremstyle{thmstyleone}%
\theoremstyle{thmstyletwo}%
\newtheorem{remark}{Remark}%

\theoremstyle{thmstylethree}%
\newtheorem{definition}{Definition}%

\begin{document}


\title[Estimating Global Identifiability Using Conditional Mutual Information in a Bayesian Framework]{Estimating Global Identifiability Using Conditional Mutual Information in a Bayesian Framework}

\author*[1]{\fnm{Sahil} \sur{Bhola}}\email{sbhola@umich.edu}

\author[1]{\fnm{Karthik} \sur{Duraisamy}}\email{kdur@umich.edu}

\affil[1]{\orgdiv{Department of Aerospace Engineering}, \orgname{University of Michigan}, \orgaddress{\city{Ann Arbor}, \postcode{48109}, \state{Michigan}, \country{USA}}}

\abstract{
A novel information-theoretic approach is proposed to assess the global practical identifiability of Bayesian statistical models.
Based on the concept of conditional mutual information, an estimate of information gained for each model parameter is used to quantify the identifiability with practical considerations.
No assumptions are made about the structure of the statistical model or the prior distribution while constructing the estimator.
The estimator has the following notable advantages: first, no controlled experiment or data is required to conduct the practical identifiability analysis; second, unlike popular variance-based global sensitivity analysis methods, different forms of uncertainties, such as model-form, parameter, or measurement can be taken into account; third, the identifiability analysis is global, and therefore independent of a realization of the parameters.
If an individual parameter has low identifiability, it can belong to an identifiable subset such that parameters within the subset have a functional relationship and thus have a combined effect on the statistical model.
The practical identifiability framework is extended to highlight the dependencies between parameter pairs that emerge a posteriori to find identifiable parameter subsets.
The applicability of the proposed approach is demonstrated using a linear Gaussian model and a non-linear methane-air reduced kinetics model.
It is shown that by examining the information gained for each model parameter along with its dependencies with other parameters, a subset of parameters that can be estimated with high posterior certainty can be found.
}



\maketitle
\section{Introduction}
\label{sec:intro}
With the growth in computational capabilities, statistical models are becoming increasingly complex to make predictions under various design conditions.
These models often contain uncertain parameters which must be estimated using data obtained from controlled experiments.
While methods for parameter estimation have matured significantly, there remain notable challenges for a statistical model and estimated parameters to be considered reliable.
One such challenge is the \textit{practical identifiability} of model parameters which is defined as the possibility of
estimating each parameter with high confidence given
different forms of uncertainties, such as parameter, model-form, or measurement are present~\citep{Bellman1970, Cobelli1980, Paulino1994}.
Low practical identifiability of the statistical model can lead to an ill-posed estimation problem which becomes a critical issue when the parameters have a physical interpretation and decisions are to be made using their estimated values~\citep{Raue2013, Lam2022}.
Further, such identifiability deficit can also lead to an unreliable model prediction, and therefore such statistical models are not suitable for practical applications~\citep{Ramancha2022, Deussen2022}.
Therefore, for a reliable parameter estimation process and model prediction, it is of significant interest that the practical identifiability is evaluated before any controlled experiment or parameter estimation studies are conducted~\citep{Ran2017, Qian2020}.

In frequentist statistics, the problem of practical identifiability is to examine the possibility of \textit{unique} estimation of model parameters $\theta$~\cite{Ran2017}.
Under such considerations, methods examining identifiability are broadly classified into \textit{local} and \textit{global} identifiability methods.
While the former examines the possibility that $\theta=\theta^k$ is a unique parameter estimate within its neighborhood $N(\theta^k)$ in the parameter space, the latter is concerned with the uniqueness of $\theta^k$ when considering the entire parameter space.
Local sensitivity analysis has been widely used to find parameters that produce large variability in the model response~\citep{Tomovic1965, Meissinger1966, Tortorelli1994, Saltelli2008}.
In such an analysis, parameters resulting in large variability are considered relevant and therefore assumed to be identifiable for parameter estimation.
However, the parameters associated with large model sensitivities could still have poor identifiability characteristics~\cite{Dobre2010}.
Another class of frequentist identification methods is based on the analysis of the properties of the Fisher information matrix (FIM).
\citet{Staley1970} proposed that the positive definiteness of the FIM is the necessary and sufficient condition for the parameters to be considered practically identifiable.
Similarly, \citet{Rothenberg1971} showed that the identifiability of the parameters is equivalent to the non-singularity of the FIM.
Later,~\citet{Stoica1982} and~\citet{Petersen2001} showed that models with singular FIM could also be identifiable.
\citet{Weijers1997} extended the classical FIM analysis and showed that even if an individual parameter has low identifiability it can belong to an identifiable subset, such that the subset is practically identifiable.
Parameters within such subsets have functional relationships with each other, thus resulting in a combined effect on the model response.
It has been shown that such identifiable subsets can be found by examining the condition number ($E$-criterion) and determinant ($D$-criterion), and selecting parameter pairs with the smallest condition number and largest determinant.
Following~\citet{Weijers1997},~\citet{Machado2009} considered the $D$ to $E$ ratio to examine practical identifiability and to find the identifiable subsets.
Another popular identification technique is likelihood profiling~\citep{Raue2009, Raue2011, Raue2013, Simpson2020}.
The method is based on finding the likelihood profile of a parameter by maximizing the likelihood with respect to the rest of the parameters.
Parameters for which its likelihood profile is shallow are deemed to have low practical identifiability.
In addition to evaluating practical identifiability, likelihood profiling could also be used to find functional relationships between parameters, which is helpful for model reparameterization~\citep{Eisenberg2014, Maiwald2016}.
However, due to the several re-optimizations required to obtain the likelihood profiles, the method does not scale well with parameter space and could quickly become computationally intractable.
While methods based on FIM or likelihood profiling have gained significant popularity they only examine local identifiability.
This means that the estimate of practical identifiability is dependent on $\theta^k$ for which the analysis is conducted and is only valid within its neighborhood $N(\theta^k)$.
To overcome the limitations of local identifiability, global identifiability methods using Kullback-Leibler divergence~\cite{Ran2014} and identifying functions~\cite{Ran2015} have been proposed.
However, such methods are computationally complex and not suitable for practical problems.
Moreover, since such methods are based on frequentist statistics, they are unable to account for parametric uncertainty and therefore unable to provide an honest representation of global practical identifiability.

There have been few studies examining global practical identifiability in a Bayesian framework.
Early attempts were based on global sensitivity analysis (GSA) that apportions the variability (either by derivatives or variance) of the model output due to the uncertainty in each parameter~\citep{Archer1997, Saltelli2000, Kucherenko2011, Ramancha2022}.
Unlike local sensitivity analysis, GSA-based methods simultaneously vary model parameters according to their distributions, thus providing a measure of global sensitivity that is independent of a particular parameter realization.
However, global parameter sensitivity does not guarantee global practical identifiability~\cite{Wu2019}.
\citet{Pant2015} and \citet{Capellari2017} formulated the problem of practical identifiability as gaining sufficient information about each model parameter from data.
An information-theoretic approach was used to quantify the information gained, such that larger information gain would mean larger practical identifiability.
However, assumptions about the structure of parameter-data joint distribution were made when developing the estimator.
A similar approach was used by~\citet{Ebrahimian2019} where the change in parameter uncertainty moving from the prior distribution to the estimated posterior distribution was used to quantify information gained.
\citet{Pant2018} proposed information sensitivity functions by combining information theory and sensitivity analysis to quantify information gain.
However, the joint distribution between the parameters and the data was assumed to be Gaussian.

Framed in a Bayesian setting, the information-theoretic approach to identifiability provides a natural extension to include different forms of uncertainties that are present in practical problems.
In this work, a novel estimator is developed from an information-theoretic perspective to examine the practical identifiability of a statistical model.
The expected information gained from the data for each model parameter is used as a metric to quantify practical identifiability.
In contrast to the aforementioned methods based on information theory, the proposed approach has the following novel advantages: first, the estimator for information gain can be used for an a priori analysis, that is, no data is required to evaluate practical identifiability; second, the framework can account for different forms of uncertainty, such as model-form, parameter, and measurement; third, the framework does not make assumptions about the joint distribution between the data and parameters as in the previous methods; fourth, the identifiability analysis is global, rather than being dependent on a particular realization of model parameters.
Another contribution of this work is an information-theoretic estimator to highlight dependencies between parameter pairs that emerge a posteriori, however, in an a priori manner.
Combining the knowledge about information gained about each parameter and parameter dependencies using the proposed approach, it is possible to find parameter subsets that can be estimated with high posterior certainty before any controlled experiment is performed.
Broadly, this can dramatically reduce the cost of parameter estimation, inform model-form selection or refinement, and associate a degree of reliability to the parameter estimation.

The manuscript is organized as follows.
In~\S\ref{ss:basesian_estimation} the Bayesian paradigm for parameter estimation is presented.
In~\S\ref{ss:quantifying_info_gain} differential entropy and mutual information are presented as information-theoretic tools to quantify the uncertainty associated with random variables and information gain, respectively.
In~\S\ref{ss:estimating_information_content} an a priori estimator is developed to quantify global practical identifiability in a Bayesian construct.
In~\S\ref{ss:estimating_parameter_dependence} the problem of estimating parameter dependencies is addressed.
An a priori estimator is developed to quantify parameter dependencies developed a posteriori.
The practical identifiability framework is applied to a linear Gaussian statistical model and methane-air reduced kinetics model; results are presented in~\S\ref{s:numerical_experiments}.
Concluding remarks are presented in~\S\ref{sec:conclusion}.

\section{Quantifying practical identifiability in a Bayesian setting}
\label{sec:identifiability}

In this section, we first present the Bayesian framework for parameter estimation.
Next, we utilize the concepts of differential entropy and mutual information from information theory to quantify information contained in the data about uncertain parameters of the statistical model.
Thereafter, we extend the idea of mutual information to develop an a priori estimator to quantify practical identifiability in a Bayesian setting.
While in most statistical models low practical identifiability is due to insufficient information about model parameters, it may often be the case that identifiable subsets exist.
Parameters within such subsets have functional relations and exhibit a combined effect on the statistical model, such that the subset is practically identifiable.
To find such identifiable subsets, we develop an estimator to highlight dependencies between parameter pairs that emerge a posteriori.

\subsection{Bayesian parameter inference}
\label{ss:basesian_estimation}

Consider the observation/data $y\in\mathcal{Y}$ of a physical system which is a realization of a random variable $\mathrm{Y}:\Omega\rightarrow\real{n}$ distributed as $p(y)$, where $\mathcal{Y}$ is the set of all possible realizations of the random variable.
Herein, we will use the same lower-case, upper-case, and symbol notation to represent a realization, random variable, and the set of all possible observations, respectively.
Consider another real-valued random variable $\Theta:\Omega\rightarrow\real{m}$ distributed as $p(\theta): \real{m}\rightarrow \real{+}$ which denotes the uncertain parameters of the model.
The data is assumed to be generated by the statistical model given as

\begin{equation}
    y\triangleq\fancy{F}(\theta, d)+\xi,
    \label{eqn:statistical_model}
\end{equation}
\noindent
where $\fancy{F}(\theta, d): \real{m}\times\real{\ell}\rightarrow\real{n}$ is the \textit{forward model} which maps the parameters and model inputs $d\in\real{\ell}$ to the prediction space.
For simplicity, consider the input of the model $d$ as known.
The random variable $\xi$ is the additive measurement noise or uncertainty in our measurement.
Once the observations are collected using controlled experiments, the \textit{prior} belief of the parameter distribution $p(\theta\mid d)$ can be updated to obtain the \textit{posterior} distribution $p(\theta\mid y, d)$ via the Bayes' rule

\begin{equation}
    p(\theta\mid y, d) = \frac{p(y\mid\theta, d)p(\theta\mid d)}{p(y\mid d)},
    \label{eqn:bayes_rule}
\end{equation}
where $p(y\mid \theta, d)$ is called the model likelihood and $p(y\mid d)$ is called the evidence.



\subsection{Quantifying information gain}
\label{ss:quantifying_info_gain}
Updating parameter belief from the prior to posterior in~(\ref{eqn:bayes_rule}) is associated with a gain in information from the data.
This gain can be quantified as the change in the uncertainty of the parameters $\Theta$.
As an example, consider a 1D Gaussian prior and posterior distribution such that the information gain can be quantified as a change in variance (a measure of uncertainty) of the parameter distribution.
\textit{A greater reduction in parameter uncertainty is a consequence of more information gained from the data.}

In general, the change in parameter uncertainty between the prior and posterior distributions for a given input of the model $d\in\mathcal{D}$ is defined as

\begin{equation}
    \Delta \mathscr{U}(d, y) \triangleq \mathscr{U}(p(\theta\mid d)) - \mathscr{U}(p(\theta\mid y, d)),
    \label{eqn:information_gained}
\end{equation}
\noindent
where $\mathscr{U}$ is an operator quantifying the amount of uncertainty or the lack of information for a given probability distribution.
Thus, the expected information gained about the parameters is defined as

\begin{equation}
    \Delta \mathscr{U}(d) \triangleq \mathscr{U}(p(\theta\mid d)) - \int\limits_{\mathcal{Y}}\mathscr{U}(p(\theta\mid y, d)) \,\mathrm{d}y.
    \label{eqn:expected_information_gained}
\end{equation}

One popular choice for the operator $\mathscr{U}$ is the differential entropy~\citep{Pant2015, Capellari2017, Aggarwal2021} which is defined as the average Shannon information~\cite{Shannon1948} for a given probability distribution.
Mathematically, for a continuous random variable $Z:\Omega\rightarrow\real{t}$ with distribution $p(z): \real{t}\rightarrow\real{+}$ and support $\mathcal{Z}$, the differential entropy is defined as

\begin{equation}
    H(p(z)) = H(Z) \triangleq -\int\limits_{\mathcal{Z}} p(z) \log p(z) \,\mathrm{d}z.
    \label{eqn:differential_entropy}
\end{equation}

Using differential entropy to quantify the uncertainty of a probability distribution, the change in uncertainty (or expected information gain) of $\Theta$ can be evaluated as

\begin{subequations}
\begin{align}
    \Delta\mathscr{U}(d) &= H(\Theta\mid d) - H(\Theta\mid Y,d),\\
                        &= H(\Theta\mid d) + H(Y\mid d) - H(\Theta, Y \mid d) \label{eqn:mutual_information_differential_form}, \\
                        &= -\int\limits_{\Theta}p(\theta\mid d)\log p(\theta\mid d) \,\mathrm{d}\theta + \int\limits_{\Theta, \mathcal{Y}} p(\theta, y\mid d) \log p(\theta\mid y, d) \,\mathrm{d}\theta \,\mathrm{d}y,\\
                        &= \int\limits_{\Theta, \mathcal{Y}} p(\theta, y\mid d) \log \frac{p(\theta, y\mid d)}{p(\theta\mid d)p(y\mid d)} \,\mathrm{d}\theta \,\mathrm{d}y,\\
                        &\triangleq I(\Theta;Y\mid d).
                        \label{eqn:mutual_information}
\end{align}
\end{subequations}
The quantity $I(\Theta;Y \mid d)$ is called mutual information between the random variables $\Theta$ and $Y$ given the model inputs $\mathcal{D}=d$~\cite{Cover1999}.
In the case of discrete random variables the mutual information is measured in \emph{bits}, whereas in the case of continuous variables the units are \emph{nats}.
\begin{remark}
    \label{remark:mutual_info}
    The mutual information $I(\Theta; Y\mid d)$ is always non-negative~\cite{Cover1999}.
    This means that updating the parameter belief from the prior to the posterior cannot increase parameter uncertainty.
\end{remark}
\subsection{Estimating practical identifiability}
\label{ss:estimating_information_content}
In a Bayesian framework where the parameters are treated as random variables, practical identifiability can be determined by examining information gained about each model parameter~\citep{Pant2015, Pant2018}.
Parameters for which the data is uninformative cannot be estimated with a high degree of confidence and therefore are practically unidentifiable.
While mutual information in (\ref{eqn:mutual_information}) is a useful quantity to study information gained from data about the entire parameter set, it does not apportion information gained about each parameter.
Therefore, to examine practical identifiability, we define a conditional mutual information
\begin{eqnarray}
    I(\Theta_{i};Y\mid\Theta_{\sim i}, d) &\triangleq& \expect{\Theta_{\sim i}}{I(\Theta_i;Y\mid \Theta_{\sim i}= \theta_{\sim i}, d)},
    \label{eqn:information_content}
\end{eqnarray}
where $\Theta_{\sim i}$ are all parameters except $\Theta_i$ and $\expect{\Theta_{\sim i}}{\cdot}$ denotes the expectation over $p(\theta_{\sim i} \mid d)$.
Using such conditional mutual information for practical identifiability is based on the intuition that on average high information gained about $\Theta_i$ means high practical identifiability.
We can thus present the following definitions for identifiability in a Bayesian setting.
\begin{definition}[Local Identifiability]
Given a statistical model with parameters $\Theta$, a parameter $\Theta_i\in\Theta$ is said to be locally identifiable if sufficient information is gained about it for a particular realization $\theta_{\sim i}$ of $\Theta_{\sim i}$.
\end{definition}
\begin{definition}[Global Identifiability]
    Given a statistical model with parameters $\Theta$, a parameter $\Theta_i\in\Theta$ is said to be globally identifiable if sufficient information is gained about it on average with respect to the distribution $p(\theta_{\sim i}\mid d)$.
\end{definition}
\noindent

The expectation over possible realizations of $\Theta_{\sim i}$ in (\ref{eqn:information_content}) therefore provides a statistical measure of global practical identifiability~\cite{Ran2017}.
On the contrary, evaluating (\ref{eqn:information_content}) at a fixed $\theta_{\sim i}$ will result in a local identifiability measure, which means that the information gained about $\Theta_i$ will implicitly depend on $\theta_{\sim i}$.

Typically, (\ref{eqn:information_content}) does not have a closed-form expression and must be estimated numerically.
Using the definition of differential entropy in (\ref{eqn:differential_entropy}) the conditional mutual information can be written as

\begin{subequations}
\begin{align}
    I(\Theta_i;Y\mid \Theta_{\sim i}, d) & = \int\limits_{\Theta_{i},\Theta_{\sim i}, \mathcal{Y}} p(\theta_i,\theta_{\sim i}, y\mid d) \log \frac{p(\theta_i, y \mid \theta_{\sim i}, d)}{p(\theta_i\mid\theta_{\sim i}, d)p(y\mid\theta_{\sim i}, d)} \,\mathrm{d}\theta_i \,\mathrm{d}\theta_{\sim i} \,\mathrm{d}y,\\
    &= \int\limits_{\Theta_{i},\Theta_{\sim i}, \mathcal{Y}} p(\theta_i,\theta_{\sim i}, y\mid d) \log \frac{p(y\mid\theta_i, \theta_{\sim i}, d)p(\theta_i\mid\theta_{\sim i}, d)}{p(\theta_i\mid\theta_{\sim i}, d)p(y\mid\theta_{\sim i}, d)} \,\mathrm{d}\theta_i \,\mathrm{d}\theta_{\sim i} \,\mathrm{d}y,\\
    &= \int\limits_{\Theta_{i},\Theta_{\sim i}, \mathcal{Y}} p(\theta_i,\theta_{\sim i}, y\mid d) \log \frac{p(y\mid\theta_i, \theta_{\sim i}, d)}{p(y\mid\theta_{\sim i}, d)} \,\mathrm{d}\theta_i \,\mathrm{d}\theta_{\sim i} \,\mathrm{d}y.\label{eqn:information_content_final_form}
\end{align}
\end{subequations}

\begin{remark}
In terms of differential entropy, the conditional mutual information in (\ref{eqn:information_content}) can be defined as

\begin{subequations}
\begin{align}
    I(\Theta_{i};Y\mid\Theta_{\sim i}, d) &\triangleq H(\Theta_i\mid\Theta_{\sim i}, d) - H(\Theta_i\mid \Theta_{\sim i}, Y, d),\\
                                &= H(\Theta_i, \Theta_{\sim i}\mid d) + H(\Theta_{\sim i}, Y\mid d)- H(\Theta_{\sim i}\mid d) - H(\Theta_i, \Theta_{\sim i}, Y\mid d).
    \label{eqn:information_content_differential_form}
\end{align}
\end{subequations}
In case the parameters are uncorrelated,
\begin{eqnarray}
    I(\Theta_i;Y\mid\Theta_{\sim i}, d) = H(\Theta_i\mid d) + H(\Theta_{\sim i}, Y\mid d) - H(\Theta_i, \Theta_{\sim i}, Y\mid d). \label{eqn:information_content_differential_form_uncorrelated}
\end{eqnarray}
While this formulation does not involve any conditional distributions involving the parameters or data, it requires joint distributions, namely, $p(\theta_i, \theta_{\sim i}\mid d)$, $p(\theta_{\sim i}, y\mid d)$, $p(\theta_i, \theta_{\sim i}, y\mid d)$.
Typically, such joint distributions do not have a closed-form expression and must be approximated.
\end{remark}

In the special case where $\Theta_i$ perfectly correlates with $\Theta_{\sim i}$ such that the realization of $\theta_{\sim i}$ provides sufficient information about $\theta_i$, the term inside the logarithm in (\ref{eqn:information_content_final_form}) becomes identically unity.
For such a case, the data is not informative about $\Theta_i$ and the effective parameter dimensionality $m_{\text{eff}}$ becomes less than $m$.
For a more general case, Monte-Carlo integration can be used to approximate the high dimensional integral as

\begin{equation}
    I(\Theta_{i};Y\mid \Theta_{\sim i}, d) \approx \hat{I}(\Theta_{i};Y\mid \Theta_{\sim i}, d) = \sum_{k=1}^{n_{\text{outer}}}\log \frac{p(y^{k}\mid\theta_{i}^{k}, \theta_{\sim i}^{k}, d)}{p(y^{k}\mid\theta_{\sim i}^{k}, d)},
    \label{eqn:estimated_conditional_mutual_information_parameter_data}
\end{equation}
\noindent
where $(\theta_{i}^{k},\theta_{\sim i}^{k})$ is drawn from the distribution $p(\theta_i, \theta_{\sim i}\mid d)$; $y^{k}$ is drawn from the likelihood distribution $p(y\mid\theta_{i}^k, \theta_{\sim i}^k, d)$; and $n_{\text{outer}}$ is the number of Monte-Carlo samples.
Typically, conditional evidence $p(y\mid \theta_{\sim i}, d)$ does not have a closed-form expression, and therefore $p(y^k\mid \theta_{\sim i}^k, d)$ must be numerically approximated.
One approach is to rewrite the conditional evidence $p(y^k\mid \theta_{\sim i}^k, d)$ by means of marginalization as

\begin{equation}
    p(y^{k}\mid \theta_{\sim i}^{k}, d) \triangleq \int\limits_{\Theta_{i}} p(y^{k},\theta_{i}\mid \theta_{\sim i}^{k}, d) \,\mathrm{d}\theta_{i} =  \int\limits_{\Theta_{i}} p(y^{k}\mid\theta_{i}, \theta_{\sim i}^{k}, d)p(\theta_i\mid \theta_{\sim i}^k, d) \,\mathrm{d}\theta_{i}.
    \label{eqn:inner_likelihood_pre_simplification}
\end{equation}
For simplicity, assume that the parameters are uncorrelated prior to observing the data, and are also independent of the model inputs $d$.
As a result, (\ref{eqn:inner_likelihood_pre_simplification}) can be re-written as

\begin{equation}
    p(y^{k}\mid \theta_{\sim i}^{k}, d) = \int\limits_{\Theta_{i}} p(y^{k}\mid\theta_{i}, \theta_{\sim i}^{k}, d)p(\theta_i) \,\mathrm{d}\theta_{i}.
    \label{eqn:inner_likelihood}
\end{equation}
This results in a low-dimensional integral over a univariate prior distribution $p(\theta_i)$.
Evaluating (\ref{eqn:inner_likelihood}) using the classical Monte-Carlo integration can dramatically increase the overall cost of estimating the conditional mutual information in (\ref{eqn:estimated_conditional_mutual_information_parameter_data}), especially if the likelihood evaluation is computationally expensive.
In the special case where the priors are normally distributed, this cost can be reduced by considering a $\zeta$-point Gaussian quadrature rule.
Using the quadrature approximation in (\ref{eqn:inner_likelihood}) gives

\begin{equation}
    p(y^{k}\mid \theta_{\sim i}^{k}, d) \approx \hat{p}(y^{k}\mid \theta_{\sim i}^{k}, d) = \sum_{\zeta=1}^{\zeta=n_{\text{inner}}} \Big[p(y^{k}\mid\theta_{i}^{\zeta}, \theta_{\sim i}^{k}, d)\Big]\gamma^{\zeta},
    \label{eqn:quadrature_information_content}
\end{equation}
\noindent
where $\theta_{i}^{\zeta}$ and $\gamma^{\zeta}$ are the $\zeta^{th}$ quadrature point and weight, respectively; $n_{\text{inner}}$ is the number of quadrature points.
Here, we use the Gauss-Hermite quadrature rule, which uses the $t^{th}$ order Hermite polynomial and will be exact for polynomials up to order $2t-1$~\cite{Sarkka2013}.
In a much more general case where the prior distributions can be non-Gaussian (however, can still be evaluated), the cost of estimating (\ref{eqn:inner_likelihood}) can be reduced by using importance sampling with a proposal distribution $q(\theta_i)$.
Using importance sampling we can rewrite (\ref{eqn:inner_likelihood}) as

\begin{align}
    p(y^{k}\mid \theta_{\sim i}^{k}, d) = \int\limits_{\Theta_{i}} \Big[p(y^{k}\mid\theta_{i}, \theta_{\sim i}^{k}, d)w(\theta_i)\Big] q(\theta_i) \,\mathrm{d}\theta_{i},
    \label{eqn:importance_sampling_information_content}
\end{align}
where $w(\theta_i) = p(\theta_i)/q(\theta_i)$ are the importance sampling weights.
In the case where the proposal distribution $q(\theta_i)$ is Gaussian, the quadrature rule can be applied to (\ref{eqn:importance_sampling_information_content}) as

\begin{equation}
    p(y^{k}\mid \theta_{\sim i}^{k}, d) \approx \hat{p}(y^{k}\mid \theta_{\sim i}^{k}, d) = \sum_{\zeta=1}^{\zeta=n_{\text{inner}}} \Big[p(y^{k}\mid\theta_{i}^{\zeta}, \theta_{\sim i}^{k}, d)w(\theta_i^{\zeta})\Big]\gamma^{\zeta}.
    \label{eqn:importance_sampling_quand_estimator}
\end{equation}
Combining the estimator for conditional evidence ((\ref{eqn:quadrature_information_content}) or (\ref{eqn:importance_sampling_quand_estimator})) with (\ref{eqn:estimated_conditional_mutual_information_parameter_data}) results in a biased estimator for conditional mutual information~\citep{Ryan2003, Huan2013}.
While the variance is controlled by the numerical accuracy of estimating the high-dimensional integral in (\ref{eqn:estimated_conditional_mutual_information_parameter_data}), the bias is governed by the accuracy of approximating the conditional evidence in (\ref{eqn:inner_likelihood_pre_simplification}).
This means that the variance is controlled by $n_{\text{outer}}$ Monte-Carlo samples and bias by $n_{\text{inner}}$ quadrature points.

In practice, estimating conditional evidence can become computationally expensive, especially when the variability in the output of the forward model is high with respect to $\Theta_i$ given $\Theta_{\sim i} = \theta_{\sim i}$, that is, large $\nabla_{\theta_{i}} \mathscr{F}(\theta, d)|_{\Theta_{\sim i}=\theta_{\sim i}}$.
For such statistical models, conditional evidence can become near zero such that numerical approximation by means of vanilla Monte-Carlo integration or Gaussian quadrature in (\ref{eqn:quadrature_information_content}) can be challenging~\cite{Huan2013}.
Using an estimator based on importance sampling for conditional evidence as shown in (\ref{eqn:importance_sampling_quand_estimator}) can alleviate this problem by carefully choosing the density of the proposal $q(\theta_i)$.
As an example, consider the case where the additive measurement noise $\xi$ is normally distributed as $\mathcal{N}(0, \Gamma)$ such that the likelihood of the model is distributed as $p(y\mid\theta)=\mathcal{N}(\mathscr{F}(\theta, d), \Gamma)$, and $y^{k}$ is sampled according to $\mathcal{N}(\mathscr{F}(\theta_{i}^{k}, \theta_{\sim i}^{k}, d), \Gamma)$.
In the case where model predictions have large variability with respect to the parameter $\Theta_i$ for a given $\Theta_{\sim i}=\theta_{\sim i}$ the model likelihood can become small.
For such a case, the importance-sampling-based estimator given in (\ref{eqn:importance_sampling_quand_estimator}) can be used by constructing a proposal around the sample $\theta_{i}^{k}$, such as $q(\theta_{i})=\mathcal{N}(\theta_{i}^{k}, \sigma^2_{\text{proposal}})$ where $\sigma^2_{\text{proposal}}$ is the variance of the proposal distribution.
This results in a robust estimation of conditional evidence and prevents infinite values for conditional mutual information.
Here, we consider (\ref{eqn:importance_sampling_quand_estimator}) to estimate conditional evidence.

\begin{remark}
    Assessing the practical identifiability in a Bayesian framework is dependent on the prior distribution.
    Although the framework presented in this article is entirely an a priori analysis of practical identifiability, prior selection can affect estimated identifiability.
    Prior selection in itself is an extensive area of research and is not considered a part of this work.
\end{remark}

\subsubsection{Physical interpretation of identifiability in an information-theoretic framework}
\label{sss:interpreting_identifiability}
Assessing practical identifiability using the conditional mutual information described in~\eqref{eqn:information_content} provides a relative measure of \emph{how many bits (or nats) of information is gained for a particular parameter}.
In practical applications where this information gain can vary on disparate scales, it is useful to associate a physical interpretation to identifiability.
Following~\cite{Pant2015}, consider a hypothetical direct observation statistical model given as $\psi \triangleq \theta_{i} + \Lambda$, where $\Lambda\sim\mathcal{N}(0, \sigma^2_{\Lambda})$ is the additive measurement noise.
Given this observation model, we can define an information gain equivalent variance $\fancy{C}(\Theta_i)$ as the \emph{measurement uncertainty in the direct observation model given $I(\Theta_i;\Psi) = \hat{I}(\Theta_i;Y\mid \Theta_{\sim i}, d)$}.
Large $\fancy{C}(\Theta_i)$ would mean that the information gained about $\Theta_i$ (using~\eqref{eqn:information_content}) for the statistical model~\eqref{eqn:statistical_model} would lead to higher measurement uncertainty if the parameter is observed directly.

If the prior distribution $p(\theta_i)$ can be approximated by means of an equivalent normal distribution $\mathcal{N}(\mu_{e}, \sigma^2_{e})$ then $I(\Theta_i;\Psi)$ is given as

\begin{align}
    I(\Theta_i;\Psi) \triangleq \frac{1}{2}\log \Big( 1 + \frac{\sigma^2_{e}}{\sigma^2_{\Lambda}}\Big),
\end{align}
such that

    \begin{align}
        \fancy{C}(\Theta_i) \triangleq \sigma^2_{\Lambda} = \sigma^2_{e} (\exp\{2\hat{I}(\Theta_i;Y\mid \Theta_{\sim i}, d)\} - 1)^{-1}.
    \end{align}
This information gain equivalent variance only depends on the information gained for the model parameter, and thus, can be used as a metric to compare different model parameters.

\subsection{Estimating parameter dependence}
\label{ss:estimating_parameter_dependence}
In most statistical models, unknown functional relationships or dependencies may be present between parameters such that multiple parameters have a combined effect on the statistical model.
Such parameters can form an identifiable subset where an individual parameter will exhibit low identifiability, however, the subset is collectively identifiable.
This means that the data is uninformative or weakly informative about an individual parameter within the subset, whereas it is informative about the entire subset.
As an example, consider the statistical model: $y = \theta_1\theta_2*d + \xi$ for which individually identifying $\Theta_1$ or $\Theta_2$ is not possible as they have a combined effect on the statistical model.
However, it is clear that $\Theta_1$ and $\Theta_2$ belong to an identifiable subset such that the pair $(\Theta_1, \Theta_2)$ is identifiable.
Thus, considering the statistical model given by $y = \theta_3*d + \xi$ where $\theta_3=\theta_1*\theta_2$ will have better identifiability characteristics.
For such statistical models, the traditional method of examining correlations between parameters is often insufficient, as it only reveals linear functional relations between random variables.

To highlight the parameter dependencies, consider the statistical model given in (\ref{eqn:statistical_model}) such that we are interested in examining the relations between $\Theta_i\in\Theta$ and $\Theta_j\in\Theta$ that emerge a posteriori.
While the conditional mutual information presented in~\S\ref{ss:estimating_information_content} provides information on the practical identifiability of an individual parameter, it does not provide information about dependencies developed between pairs of parameters.
To quantify such dependencies, we define a conditional mutual information between parameter pairs

\begin{equation}
    I(\Theta_i;\Theta_j\mid Y, \Theta_{\sim i, j}, d) \triangleq \expect{\Theta_{\sim i, j}}{\expect{Y}{I(\Theta_i;\Theta_j\mid Y=y, \Theta_{\sim i, j}=\theta_{\sim i, j}, d)}},
    \label{eqn:parameter_pair_conditional_mi}
\end{equation}
which evaluates the average information between the variables $\Theta_i$ and $\Theta_j$ that is obtained a posteriori.
Here, $\Theta_{\sim i, j}$ is defined as all the parameters of the statistical model except $\Theta_i$ and $\Theta_j$.

A closed-form expression for (\ref{eqn:parameter_pair_conditional_mi}) is typically not available, and therefore a numerical approximation is required.
In integral form, (\ref{eqn:parameter_pair_conditional_mi}) is given as
\begin{eqnarray}
    \begin{split}
        I(\Theta_{i};\Theta_{j}\mid Y, \Theta_{\sim i, j}, d) \triangleq  \int\limits_{{\Theta_{i}, \Theta_{j}, \Theta_{\sim i, j}}, \mathcal{Y}} p(\theta_{i}, \theta_{j}, \theta_{\sim i, j}, y\mid d) \Big[\log\big[ p(\theta_{i}, \theta_j \mid y, \theta_{\sim i, j}, d)\big]- \\
        \log\big[p(\theta_{i}\mid y, \theta_{\sim i, j}, d)p(\theta_j\mid y, \theta_{\sim i, j}, d)\big]\Big] \,\mathrm{d}\theta_{i} \,\mathrm{d}\theta_{j} \,\mathrm{d}\theta_{\sim i, j} \,\mathrm{d}y,
    \end{split}
    \label{eqn:parameter_depedence_integral_formulation}
\end{eqnarray}
where

\begin{subequations}
\begin{align}
    p(\theta_i, \theta_j\mid y, \theta_{\sim i, j}, d) &\triangleq \frac{p(y\mid\theta_i, \theta_j, \theta_{\sim i, j}, d)p(\theta_i, \theta_j, \mid\theta_{\sim i, j}, d)}{p(y\mid\theta_{\sim i, j}, d)},\label{eqn:likelihood_parameter_dependence}\\
    p(\theta_i \mid y, \theta_{\sim i, j}, d) &\triangleq \frac{p(y\mid\theta_i,\theta_{\sim i, j}, d)p(\theta_i\mid\theta_{\sim i, j}, d)}{p(y\mid\theta_{\sim i, j}, d)},\label{eqn:marginal_evidence_i_parameter_dependence}\\
    p(\theta_j \mid y, \theta_{\sim i, j}, d) &\triangleq \frac{p(y\mid\theta_j,\theta_{\sim i, j}, d)p(\theta_j\mid\theta_{\sim i, j}, d)}{p(y\mid\theta_{\sim i, j}, d)},\label{eqn:marginal_evidence_j_parameter_dependence}
\end{align}
\end{subequations}
via Bayes' theorem.
\begin{remark}
In terms of differential entropy, the  conditional mutual information in (\ref{eqn:parameter_pair_conditional_mi}) can be defined as

\begin{subequations}
\begin{align}
     I(\Theta_i;\Theta_j\mid Y, \Theta_{\sim i, j}, d) &\triangleq H(\Theta_i\mid Y, \Theta_{\sim i, j}, d) - H(\Theta_i\mid \Theta_j, Y, \Theta_{\sim i, j}, d),\\
    \begin{split}
     &= H(\Theta_i, \Theta_{\sim i, j}, Y\mid d) + H(\Theta_j, \Theta_{\sim i, j}, Y\mid d) \\ &\qquad-H(\Theta_{\sim i,j}, Y\mid d) - H(\Theta_i, \Theta_j,\Theta_{\sim i, j}, Y\mid d).\end{split}
\label{eqn:parameter_dependence_differential_form}
\end{align}
\end{subequations}

Such a formulation requires evaluating joint distributions, namely, $p(\theta_i, \theta_{\sim i, j}, y\mid d)$, $p(\theta_j, \theta_{\sim i, j}, y\mid d)$, $p(\theta_{\sim i, j}, y\mid d)$, and $p(\theta_i, \theta_j, \theta_{\sim i, j}, y\mid d)$.
Typically, such joint distributions do not have a closed-form expression and must be approximated.
\end{remark}
For the sake of illustration, assume that the parameters are uncorrelated with each other \textit{prior to observing the data}.
As a consequence of this assumption, any relations developed between $\Theta_i$ and $\Theta_j$ are discovered solely from data.
Furthermore, it is also reasonable to assume that prior knowledge of the parameters is independent of the input of the model $d$.
Substituting (\ref{eqn:likelihood_parameter_dependence}) through (\ref{eqn:marginal_evidence_j_parameter_dependence}) into (\ref{eqn:parameter_depedence_integral_formulation}) we obtain

\begin{align}
    \begin{split}
        I(\Theta_{i};\Theta_{j}\mid Y, \Theta_{\sim i, j}, d)  = \int\limits_{\Theta_{i}, \Theta_{j}, \Theta_{\sim i, j}, \mathcal{Y}} &p(\theta_{i}, \theta_{j}, \theta_{\sim i, j}, y\mid d) \Big [\log \big[p(y\mid \theta_{i}, \theta_j,\theta_{\sim i, j} d)\big] \\
        &+ \log \big[p(y\mid\theta_{\sim i, j}, d)\big]\\
        &- \log\big[p(y\mid\theta_{i}, \theta_{\sim i, j}, d)\big]\\
        &- \log\big[p(y\mid\theta_j,\theta_{\sim i,j}, d)\big]\Big] \,\mathrm{d}\theta_{i} \,\mathrm{d}\theta_{j} \,\mathrm{d}\theta_{\sim i, j} \,\mathrm{d}y.
    \label{eqn:parameter_depedence_integral_formulation_final_form}
    \end{split}
\end{align}


Similar to~\S\ref{ss:estimating_information_content} we can estimate the conditional mutual information in (\ref{eqn:parameter_depedence_integral_formulation_final_form}) using Monte-Carlo integration as $\hat{I}(\Theta_i;\Theta_j\mid Y, \Theta_{\sim i, j}, d) \approx I(\Theta_i;\Theta_j\mid Y, \Theta_{\sim i, j}, d)$ where

\begin{equation}
    \begin{split}
         \hat{I}(\Theta_i;\Theta_j\mid Y, \Theta_{\sim i, j}, d) = \sum_{k=1}^{k=n_{\text{outer}}}\log \frac{p(y^{k}\mid \theta_{i}^{k}, \theta_j^{k},\theta_{\sim i, j}^{k}, d)p(y^{k}\mid\theta_{\sim i, j}^{k}, d)}{p(y^{k}\mid\theta_{i}^{k}, \theta_{\sim i, j}^{k}, d)p(y^{k}\mid\theta_j^{k},\theta_{\sim i,j}^{k}, d)},
    \end{split}
    \label{eqn:estimated_conditional_mutual_information_parameter_pair}
\end{equation}
where $\theta_i^{k}$, $\theta_j^{k}$, and $\theta_{\sim i, j}^{k}$ are drawn from the prior distributions $p(\theta_i)$, $p(\theta_j)$, and $p(\theta_{\sim i, j})$, respectively; $y^{k}$ is drawn from the likelihood distribution $p(y\mid\theta_i^k, \theta_j^k, \theta_{\sim i, j}^k, d)$.
The conditional evidence in (\ref{eqn:estimated_conditional_mutual_information_parameter_pair}) can be obtained by means of marginalization

\begin{subequations}
 \begin{align}
     p(y^{k}\mid\theta_{\sim i, j}^{k}, d)  &\triangleq \int\limits_{\Theta_i, \Theta_j} p(y^{k}\mid \theta_i, \theta_j, \theta_{\sim i, j}^{k}, d)p(\theta_i, \theta_j) \,\mathrm{d}\theta_i \,\mathrm{d}\theta_j,\label{eqn:conditional_marginal_parameter_depedence_1}\\
     p(y^{k}\mid\theta_{i}^{k}, \theta_{\sim i, j}^{k}, d)  &\triangleq \int\limits_{\Theta_j} p(y^{k}\mid \theta_j, \theta_{i}^{k}, \theta_{\sim i, j}^{k}, d)p(\theta_j) \,\mathrm{d}\theta_j,\label{eqn:conditional_marginal_parameter_depedence_2}\\
     p(y^{k}\mid\theta_{j}^{k}, \theta_{\sim i, j}^{k}, d)  &\triangleq \int\limits_{\Theta_i} p(y^{k}\mid \theta_i, \theta_{j}^{k}, \theta_{\sim i, j}^{k}, d)p(\theta_i) \,\mathrm{d}\theta_i.\label{eqn:conditional_marginal_parameter_depedence_3}
 \end{align}
\end{subequations}
Similar to~\S\ref{ss:estimating_information_content} the conditional evidence in (\ref{eqn:conditional_marginal_parameter_depedence_1}) through (\ref{eqn:conditional_marginal_parameter_depedence_3}) can be efficiently estimated using importance sampling along with Gaussian quadrature rules. However, it should be noted that (\ref{eqn:conditional_marginal_parameter_depedence_1}) is an integral over a two-dimensional space, and therefore requires $n_{\text{inner}}^{2}$ quadrature points.

\section{Numerical experiments}
\label{s:numerical_experiments}
This section presents numerical experiments to validate the information-theoretic approach to examine practical identifiability.
The estimate obtained for global identifiability is compared with the variance-based global sensitivity analysis by means of first-order Sobol indices (\suppinfo[\S\ref{sec:sobol_index}]).
First, a linear Gaussian statistical model is considered for which practical identifiability can be analytically examined through the proposed information-theoretic approach.
This model is computationally efficient and is therefore ideal for conducting estimator convergence studies.
Next, the practical identifiability of a reduced kinetics model for methane-air combustion is considered.
Reduced kinetics models are widely used in the numerical analysis of chemically reactive flows since embedding detailed chemistry of combustion is often infeasible.
Such reduced kinetic models are often parameterized such that constructing models with practically identifiable parameters is desirable to improve confidence in the model prediction.

\subsection{Application to a linear Gaussian model}
\label{sec:linear_gaussian}
The identifiability framework is now applied to a linear Gaussian problem for which closed-form expressions are available for the conditional mutual information in (\ref{eqn:information_content}) and (\ref{eqn:parameter_pair_conditional_mi}) (\suppinfo[\S\ref{sec:linear_gaussian_app}]).
Consider the statistical model
\begin{eqnarray}
    y = \mathscr{F}(\theta, d)+ \xi\quad;\xi\sim\mathcal{N}(0, \Gamma),
    \label{eqn:linear_gaussian_model}
\end{eqnarray}
where $\mathscr{F}(\theta, d) = \mathbf{A}\theta$ and $\mathbf{A}\in\real{n\times m}$ is called the feature matrix.
The prior distribution is given by $p(\theta) = \mathcal{N}(\mu_{\Theta}, \Sigma_{\Theta})$ where $\mu_{\Theta} \in \real{m}$ and $\Sigma_{\Theta}\in\real{m\times m}$.
Model likelihood is therefore given by $p(y\mid \theta) = \mathcal{N}(\mathbf{A}\theta, \Gamma)$ where $\Gamma\in\real{n\times n}$.
Here, $\mu_{\Theta}, \Sigma_{\Theta}$, and $\Gamma$ are all constants and are considered known.
The evidence distribution for this model is given by $p(y) \triangleq \mathcal{N}(\mu_{Y}, \Sigma_{Y}) =  \mathcal{N}(\mathbf{A}\mu_{\Theta}, \mathbf{A}\Sigma_{\Theta}\mathbf{A}^{T}+\Gamma)$,
such that no model-form error exists.
Consider a feature matrix

\begin{equation}
    \mathbf{A} = \begin{pmatrix}
        d_{1} & d_{1}^{2} &\dots & d_{1}^{m} \\
        d_{2} & d_{2}^{2} &\dots & d_{2}^{m} \\
        \vdots & \vdots &\ddots & \vdots \\
        d_{n} & d_{n}^{2} &\dots & d_{n}^{m} \\

                    \end{pmatrix},
\end{equation}
where $d_{i\mid_{i=1}^n}$ are $n$ linearly-spaced points in an interval $[-1, 1]$, and $m=3$, which means that the statistical model has 3 uncertain parameters.
Assume an uncorrelated measurement noise $\Gamma=\sigma_{\xi}^2\mathbb{I}$ with $\sigma_{\xi}^2 = 0.1$.
For the purpose of parameter estimation, synthetic data is generated using (\ref{eqn:linear_gaussian_model}) assuming $\theta^{*}=[1, 2, 3]^T$ and $n=100$.
\begin{figure}[ht]
    \centering
    \includegraphics[scale=0.2]{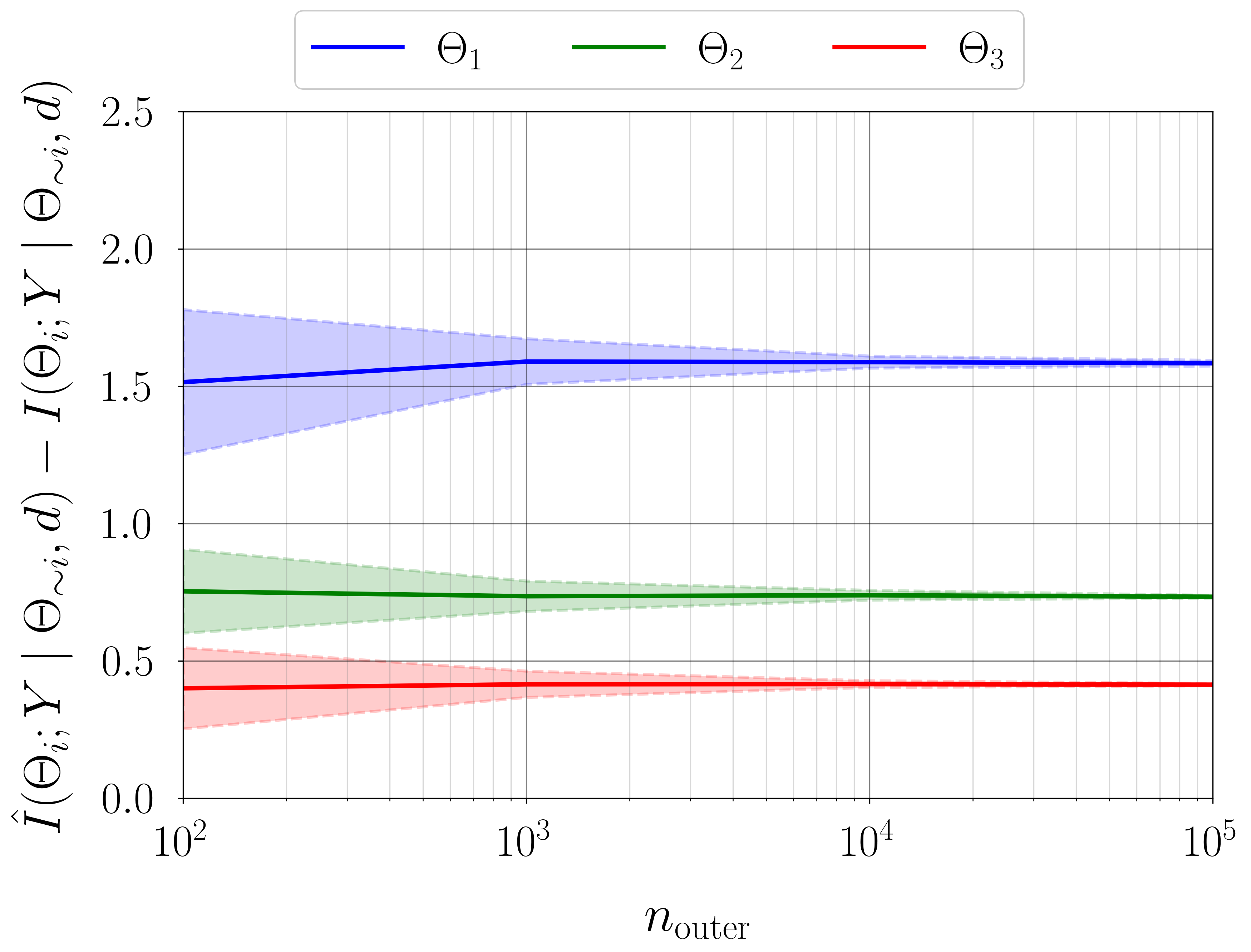}
    \includegraphics[scale=0.2]{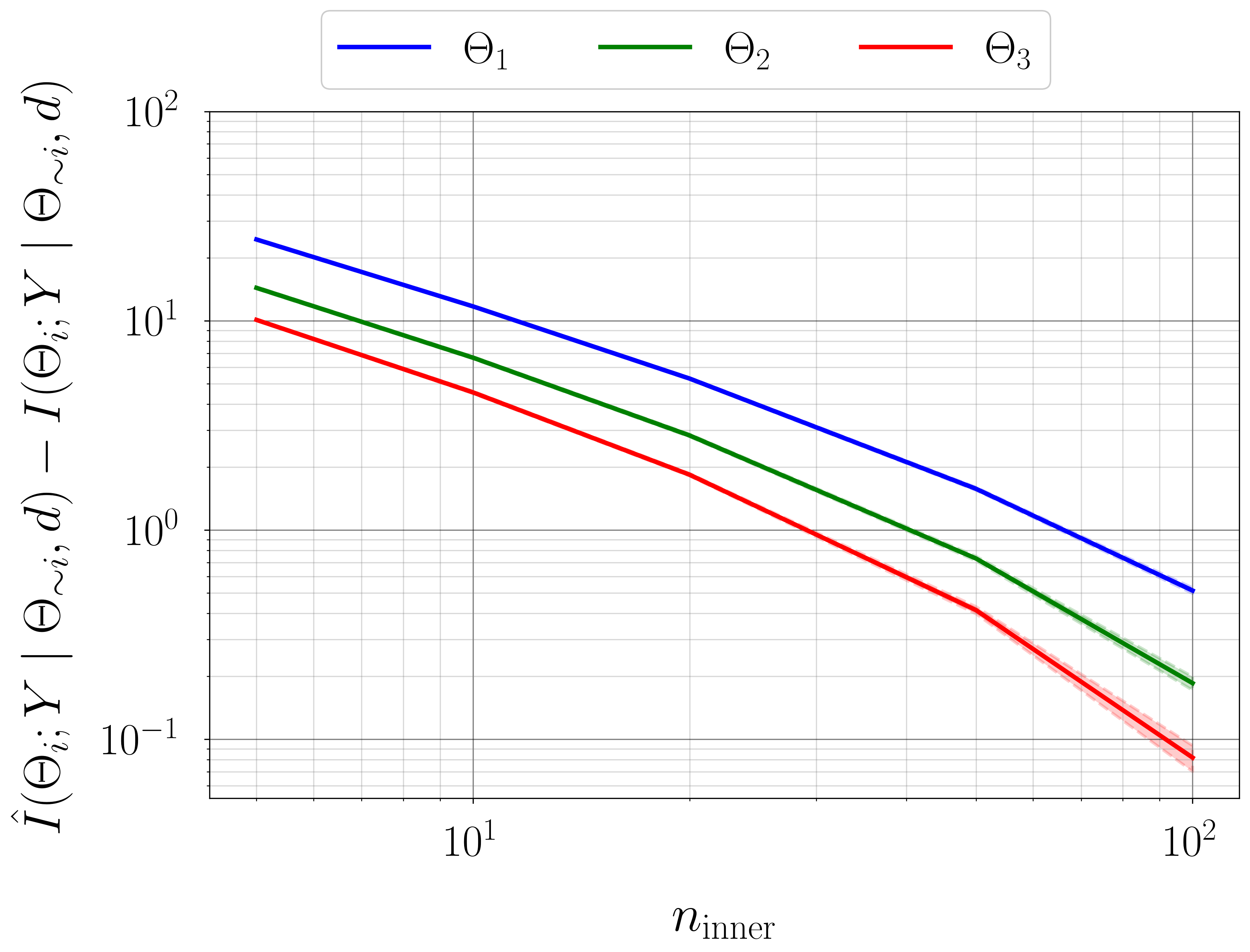}
    \caption{Convergence of the variance in practical identifiability estimator of a linear Gaussian statistical model (left); the number of quadrature points $n_{\text{inner}} = 50$ is considered and the number of Monte-Carlo integration samples $n_{\text{outer}}$ is varied.
    Bias convergence for practical identifiability estimator in case of a linear Gaussian statistical model (right); $n_{\text{outer}} = 10^{4}$ is considered and $n_{\text{inner}}$ is varied.
    For a given $n_{\text{inner}}$ increasing $n_{\text{outer}}$ reduces the variance, whereas increasing $n_{\text{inner}}$ for a given $n_{\text{outer}}$ decreases the bias in the estimate.}
    \label{fig:individual_mi_convergence}
\end{figure}

\begin{figure}[ht]
    \centering
    \includegraphics[scale=0.2]{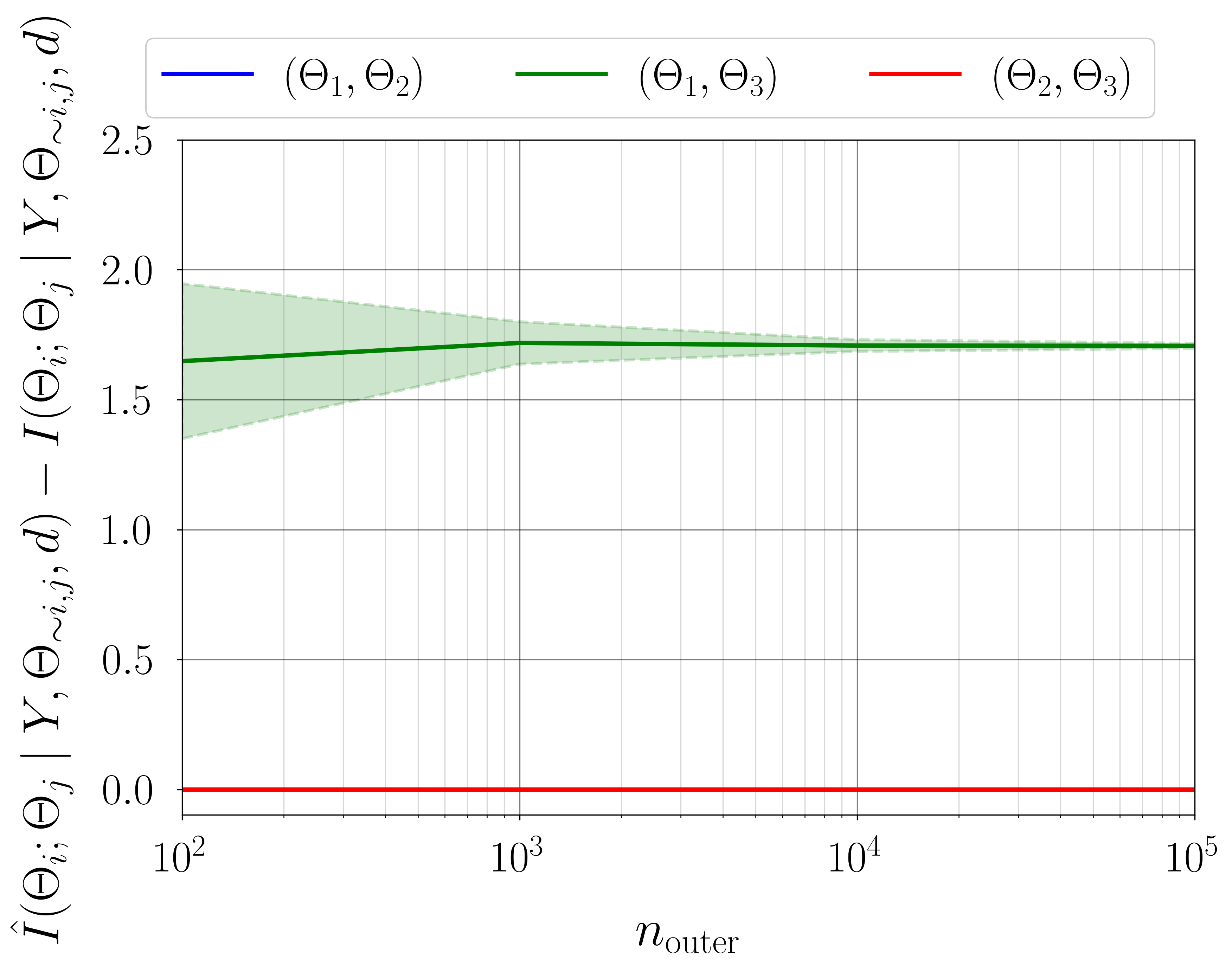}
    \includegraphics[scale=0.2]{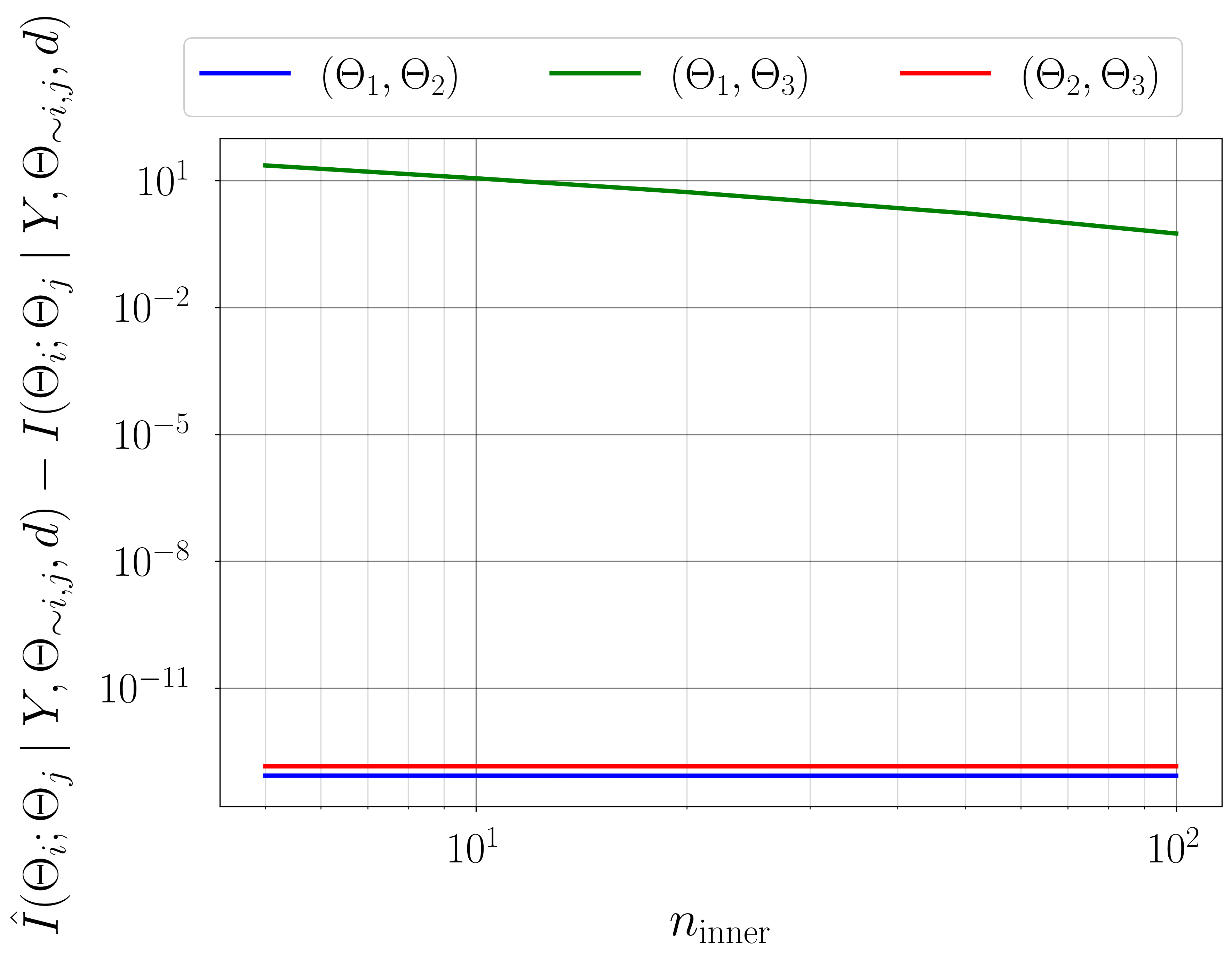}
    \caption{Convergence of the variance in practical identifiability for estimating parameter dependencies of a linear Gaussian statistical model (left); the number of quadrature points $n_{\text{inner}} = 50$ is considered and the number of Monte-Carlo integration samples $n_{\text{outer}}$ is varied.
    Bias convergence for estimating parameter dependencies in the case of a linear Gaussian statistical model (right); $n_{\text{outer}} = 10^{4}$ is considered and $n_{\text{inner}}$ is varied.
    For a given $n_{\text{inner}}$ increasing $n_{\text{outer}}$ reduces the variance, whereas increasing $n_{\text{inner}}$ for a given $n_{\text{outer}}$ decreases the bias in estimating the parameter dependencies.}
    \label{fig:pair_mi_convergence}
\end{figure}

\subsubsection{Parameter identifiability}
The goal of the framework developed in~\S\ref{ss:estimating_information_content} is to assess the practical identifiability of the statistical model in (\ref{eqn:linear_gaussian_model}) before any controlled experiment is conducted.
Consider $\mu_{\Theta} = \mathbf{0}$ and $\Sigma_{\Theta} = \mathbb{I}$.
Using such an uncorrelated prior distribution for the identifiability study ensures that the information obtained is only due to the observation of the data (as discussed in~\S\ref{ss:estimating_parameter_dependence}).
Using historical parameter estimates can improve the prior (Remark~\ref{remark:mutual_info}) which can affect the identifiability analysis.
However, we have not considered any such prior refinement.

Figure~\ref{fig:individual_mi_convergence} illustrates the convergence of error in estimating information gain for each parameter using the estimator developed in~\S\ref{ss:estimating_information_content}.
As expected, for a fixed number of quadrature points, increasing the number of Monte-Carlo integration points decreases the variance in estimation.
However, for a fixed $n_{\text{outer}}$ increasing the number of quadrature points reduces the bias in the estimate.
Figure~\ref{fig:pair_mi_convergence} illustrates the variance and bias convergence of error in estimating parameter dependencies as described in~\S\ref{ss:estimating_parameter_dependence}.
As expected and observed, the variance in error is controlled by the accuracy of Monte-Carlo integration, that is, by $n_{\text{outer}}$, and the bias is controlled by the quadrature approximation, that is, through $n_{\text{inner}}$.
\begin{figure}[ht]
    \centering
    \includegraphics[scale=0.28]{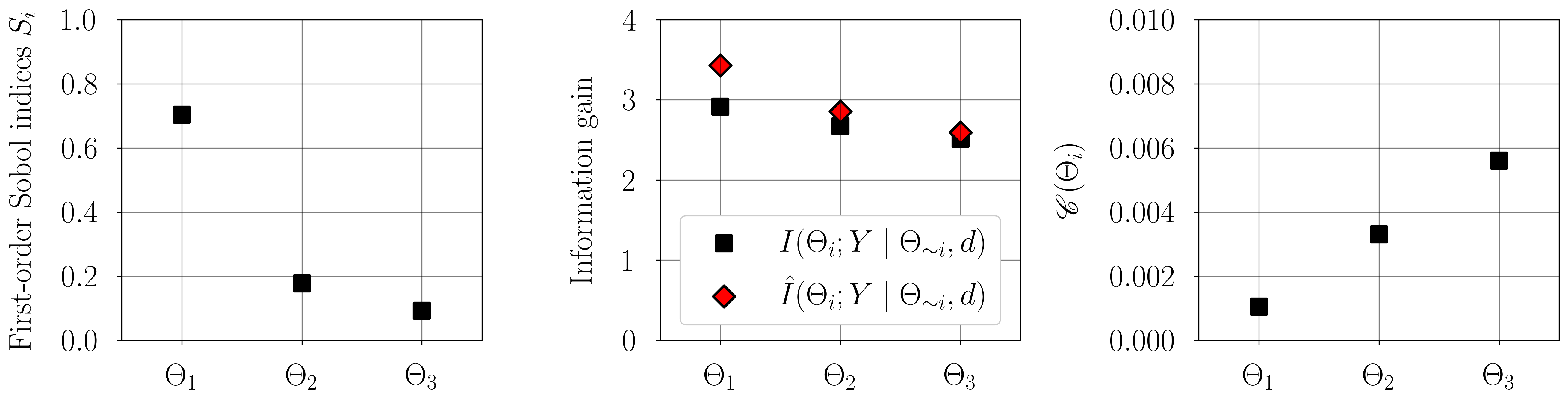}
    \caption{First-order Sobol indices (left), information gain (center), and information gain equivalent variance $\fancy{C}(\Theta_i)$ (right) for linear Gaussian model.
    Sobol indices show that the output of the statistical model has the largest variability due to uncertainty in $\Theta_1$, followed by $\Theta_2$ and $\Theta_3$.
    Variable $\Theta_{1}$ exhibits the largest gain in information and therefore the highest practical identifiability, followed by $\Theta_2$ and then $\Theta_3$.
    For a direct observation model, the variable $\Theta_1$ has the lowest measurement uncertainty, followed by $\Theta_2$ and $\Theta_3$.
    }
    \label{fig:linear_gaussian/iden_vs_sobol}
\end{figure}

\begin{figure}[ht]
    \centering
    \includegraphics[scale=0.30]{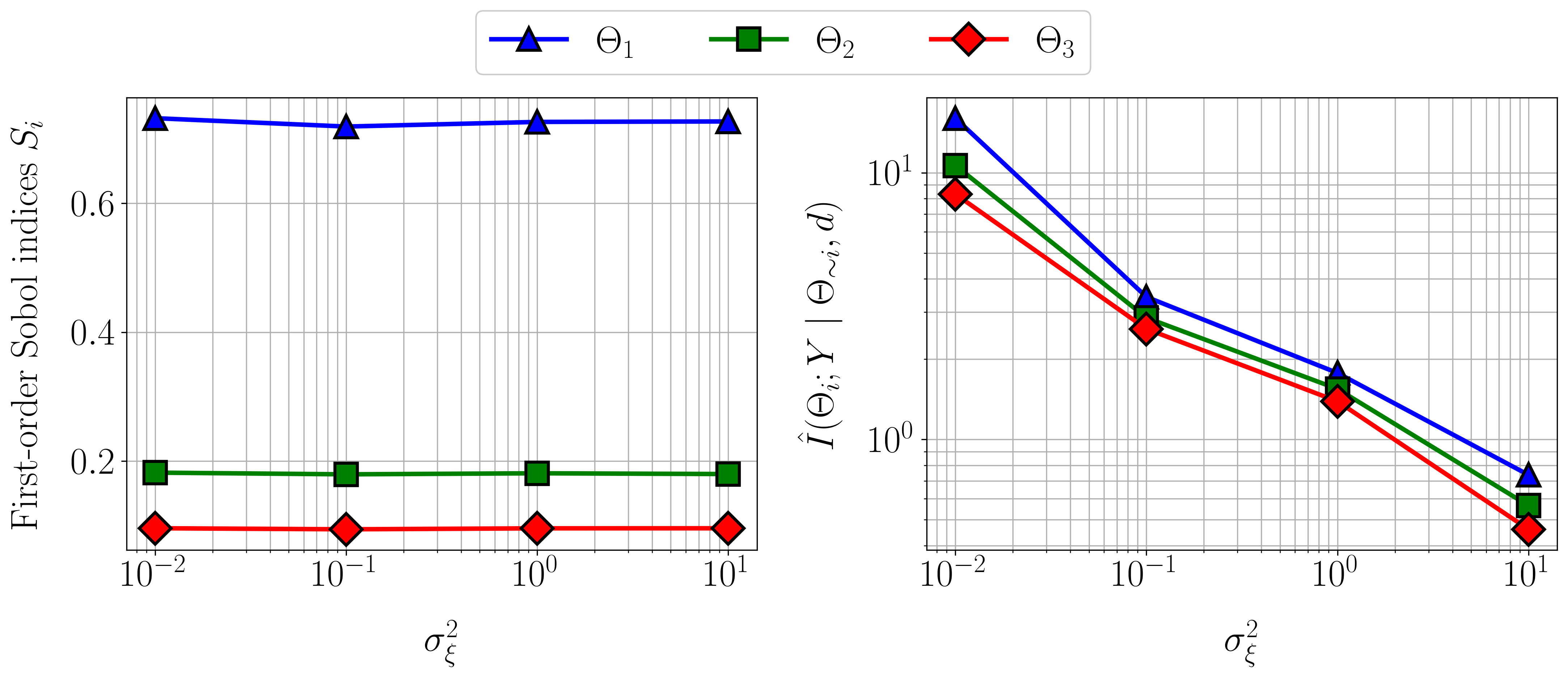}
    \caption{First-order Sobol indices (left) and estimated information gain (right) vs. measurement noise variance $\sigma_{\xi}^2$ for linear Gaussian model.
    Increasing measurement noise covariance does not affect the variability of the output with respect to the parameters and therefore the first-order Sobol indices remain unchanged.
    However, the information gain decreases with increasing measurement noise covariance.
    }
    \label{fig:linear_gaussian/info_gain_sobol_vs_noise_cov}
\end{figure}

The first-order Sobol indices, estimated information gain, and information gain equivalent variance $\fancy{C}(\Theta_i)$ are shown in figure~\ref{fig:linear_gaussian/iden_vs_sobol}.
The estimated first-order Sobol indices (\suppinfo[figure~\ref{fig:linear_gaussian_sobol_convergence}] for convergence study) show that the considered linear Gaussian forward model has the largest output variability due to uncertainty in $\Theta_1$, followed by $\Theta_2$ and $\Theta_3$.
This implies that the forward model is most sensitive to the parameter $\theta_1$, followed by $\theta_2$ and then $\theta_3$.
This is not surprising since $d_{i\mid_{i=1}^n}$ are points in the interval $[-1, 1]$.
Thus, according to the first-order Sobol indices, the relevance of the parameters follows the order: $\theta_1$, $\theta_2$, and $\theta_3$.
The estimated information gained agrees well with the truth.
Further, the obtained trend suggests that the data is most informative about $\Theta_1$, followed by $\Theta_2$, and then $\Theta_3$.
As discussed in~\S\ref{ss:estimating_information_content}, practical identifiability follows the same trend.
Furthermore, as suggested by~\citet{Wu2019}, it can be seen that parameters with good identifiability characteristics also exhibit high model sensitivity.
Using the hypothetical direct observation model described in~\S\ref{sss:interpreting_identifiability}, the smallest measurement uncertainty is obtained for the variable $\Theta_1$, followed by $\Theta_2$ and $\Theta_3$.
That is, parameters with high practical identifiability are associated with low measurement uncertainty in a direct observation model.

Figure~\ref{fig:linear_gaussian/info_gain_sobol_vs_noise_cov} illustrates the variability of the first-order Sobol indices and the estimated information gain with measurement noise variance $\sigma_{\xi}^2$.
The first-order Sobol indices only account for the parameter uncertainty, and therefore remain unchanged with an increase in measurement noise.
However, the estimated information gain and thereby the practical identifiability decreases with measurement noise.
This observation corroborates the intuition that large measurement uncertainty will lead to large uncertainty in the parameter estimation.

\begin{figure}[ht]
    \centering
    \includegraphics[scale=0.049]{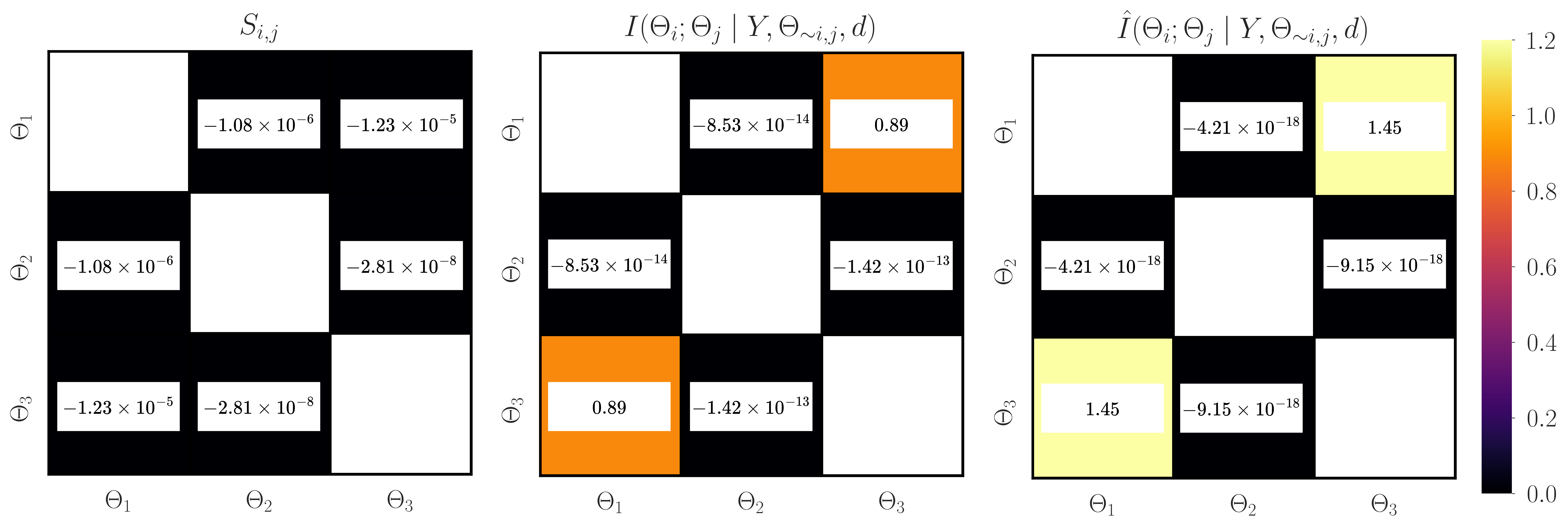}
    \caption{Second-order Sobol indices (left), true parameter dependencies (center), and estimated parameter dependencies (right) for linear Gaussian model.
    The second-order Sobol indices show negligible interactions between parameter pairs.
    The obtained estimate of parameter dependency agrees well with their true values; the trend is preserved.
    $\Theta_1$ and $\Theta_3$ have the largest dependency on one another, and therefore are expected to have a combined effect on the output of the statistical model.}
    \label{fig:linear_gaussian/true_vs_estimated_dependencies}
\end{figure}

Figure~\ref{fig:linear_gaussian/true_vs_estimated_dependencies} shows the second-order Sobol indices and the true and estimated dependencies between the parameter pairs for the linear Gaussian model.
Examining the second-order Sobol indices (\suppinfo[figure~\ref{fig:linear_gaussian_sobol_convergence}] for convergence study) shows that there are negligible interactions between parameter pairs.
Estimated parameter dependencies agree well with the truth; the trend is preserved.
The bias observed is due to the error in approximating the conditional evidence, as shown in figure~\ref{fig:pair_mi_convergence}.
It can be clearly seen that the parameters $\Theta_1$ and $\Theta_3$ have high dependencies.
This means that these parameters compensate for one another such that they will have a combined effect on the output of the statistical model.
These parameters are associated with the features $d_{i\mid_{i=1}^n}$ and $d_{i\mid_{i=1}^n}^3$ which, in fact, have a similar effect on the statistical model for $d_{i\mid_{i=1}^n}\in[-1, 1]$.
This observation also shows that the low practical identifiability of $\Theta_3$ is mainly due to the underlying dependency with $\Theta_1$ such that the pair $(\Theta_1, \Theta_3)$ has a combined effect in the statistical model.

\begin{figure}[ht]
    \centering
    \includegraphics[scale=0.31]{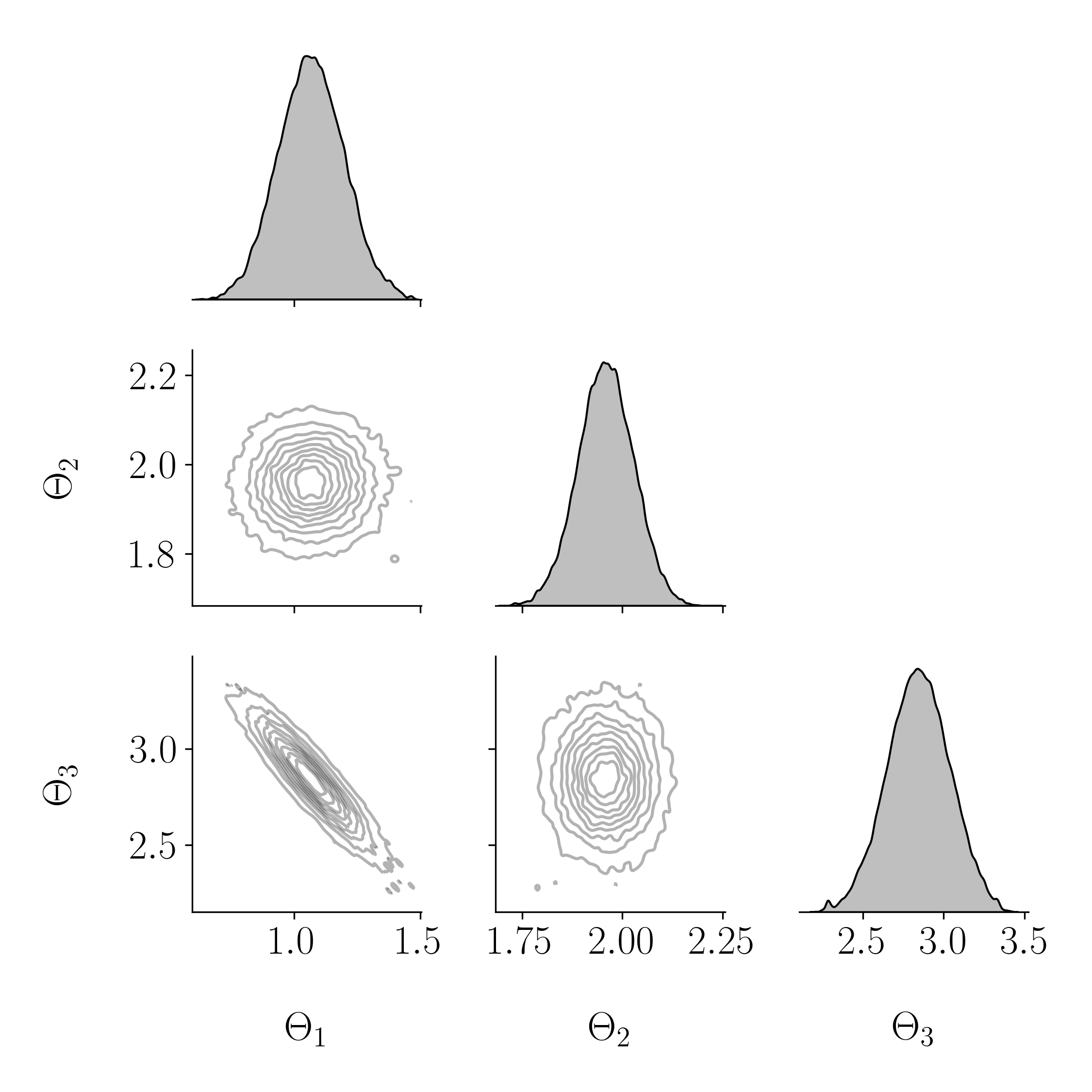}
    \includegraphics[scale=0.25]{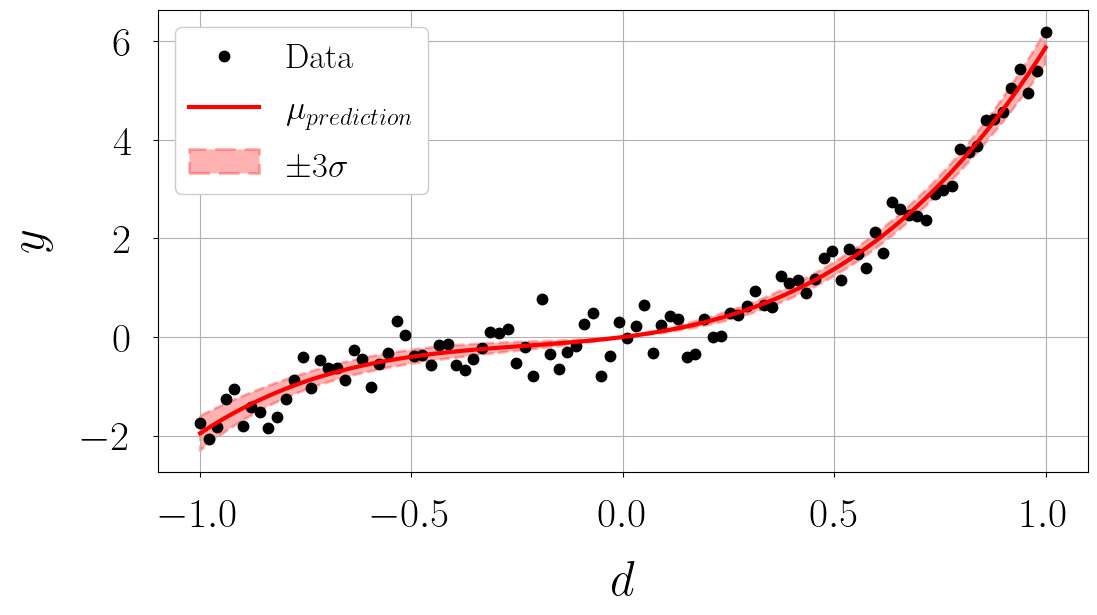}
    \caption{Correlation plot for samples obtained from the true posterior distribution (left) and the obtained aggregate posterior prediction (right) for linear Gaussian model.
    A negative correlation is observed between $\Theta_1$ and $\Theta_3$, whereas $\Theta_2$ is uncorrelated from other parameters.
    Aggregate posterior prediction agrees well with the data and exhibits high certainty.
    }
    \label{fig:linear_gaussian/posterior_prediciton}
\end{figure}
\subsubsection{Parameter estimation}
For the linear Gaussian model, the joint distribution $p(\theta, y)$ can be written as

\begin{equation}
    p(\theta, y) \triangleq \mathcal{N}(\mu_{\Theta, Y}, \Sigma_{\Theta, Y}) = \mathcal{N} \begin{pmatrix} \begin{bmatrix}\mu_{\Theta} \\ \mu_{Y}\end{bmatrix},
        \begin{bmatrix}\Sigma_{\Theta} & \Sigma_{\Theta}\mathbf{A}^{T} \\ \mathbf{A}\Sigma_{\Theta} & \Sigma_{Y} \end{bmatrix}
\end{pmatrix},
\label{eqn:total_joint}
\end{equation}
such that the analytical posterior distribution is given as $p(\theta\mid y) = \mathcal{N}(\mu_{\Theta_{post}}, \Sigma_{\Theta_{post}})$,
where $\mu_{\Theta_{post}} = \mu_{\Theta} + \Sigma_{\Theta}\mathbf{A}^{T}\Sigma_{Y}^{-1}(y-\mu_Y)$ and $\Sigma_{\Theta_{post}} = \Sigma_{\Theta} - \Sigma_{\Theta}\mathbf{A}^{T}\Sigma_{Y}^{-1}A\Sigma_{\Theta}$ using Gaussian conditioning.

Samples from the posterior distribution and the aggregate posterior prediction are shown in figure~\ref{fig:linear_gaussian/posterior_prediciton}.
Variables $\Theta_1$ and $\Theta_3$ have a negative correlation, whereas $\Theta_2$ is uncorrelated with other parameters.
This means that the parameter variables $\Theta_1$ and $\Theta_3$ have (linear) dependencies on each other, and $\Theta_2$ does not have such dependencies.
These dependencies were suggested during the a priori analysis conducted on the statistical model as illustrated in figure~\ref{fig:linear_gaussian/true_vs_estimated_dependencies}.
Aggregate posterior prediction agrees well with the data and exhibits high certainty.

\begin{figure}[ht]
    \centering
    \includegraphics[scale=0.31]{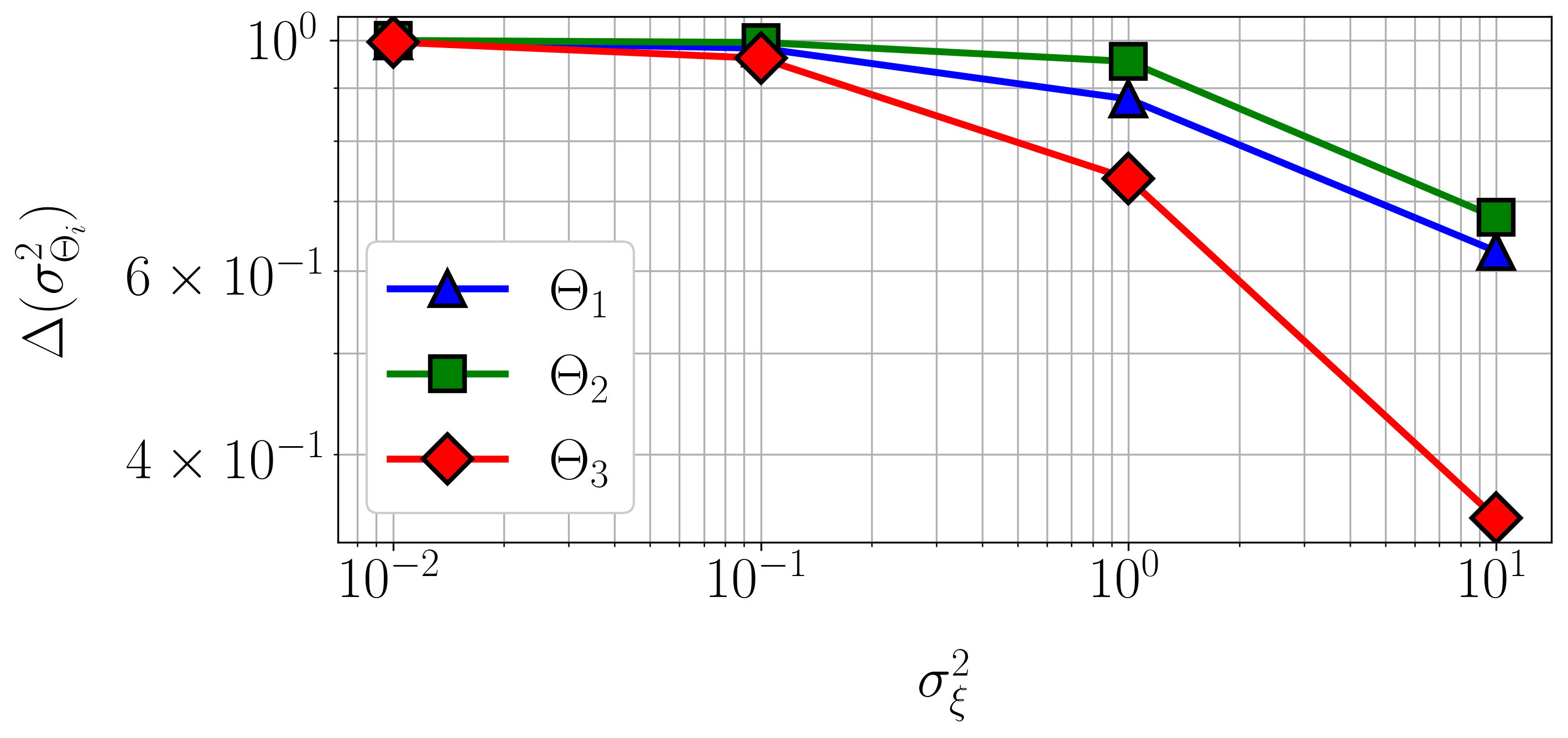}
    \caption{Change in parameter variance $\Delta(\sigma^2_{\Theta_i})$ vs. measurement noise covariance $\sigma_{\xi}^2$ for linear Gaussian model.
    Increasing measurement noise results in a smaller change in parameter variance from the prior to the posterior.
    Largest reduction in variance is observed for $\Theta_2$, followed by $\Theta_{1}$ and $\Theta_{3}$.
    }
    \label{fig:linear_gaussian/posterior_cov_vs_noise}
\end{figure}

Figure~\ref{fig:linear_gaussian/posterior_cov_vs_noise} illustrates the change in variance of the parameter $\Theta_i$ defined as $\Delta(\sigma^2_{\Theta_i}) \triangleq \sigma^2_{\Theta_i} - \sigma^2_{\Theta_{i, post}}$ versus $\sigma_{\xi}^2$.
Parameter $\Theta_2$ exhibits the smallest posterior uncertainty, followed by $\Theta_1$ and $\Theta_3$ for all $\sigma_{\xi}^2$.
While $\Theta_1$ has the largest estimated information gain (figure~\ref{fig:linear_gaussian/iden_vs_sobol} (center)), it exhibits dependencies with $\Theta_3$ (figure~\ref{fig:linear_gaussian/true_vs_estimated_dependencies} (right)), thereby resulting in larger posterior uncertainty in comparison to $\Theta_2$.
In practical applications, where model selection or parameter selection is critical, examining the information gain and parameter dependencies can therefore aid in finding parameters that can be estimated with high certainty.
Increasing the measurement noise results in a smaller change in parameter variance, that is, the parameters exhibit larger posterior uncertainty.
This is also shown by the variation of estimated information gain with measurement noise (figure~\ref{fig:linear_gaussian/info_gain_sobol_vs_noise_cov} (right)).
On the contrary, the first-order Sobol indices remain unchanged with measurement noise (figure~\ref{fig:linear_gaussian/info_gain_sobol_vs_noise_cov} (left)).



\subsection{Application to methane chemical kinetics}
Accurate characterization of chemical kinetics is critical in the numerical prediction of reacting flows.
Although there have been significant advancements in computational architectures and numerical methods, embedding the full chemical kinetics in numerical simulations is almost always infeasible.
This is primarily because of the high-dimensional coupled ordinary differential equations that have to be solved to obtain concentrations of a large number of involved species.
As a result, significant efforts have been made to develop reduced chemical kinetics models that seek to capture features such as ignition delay, adiabatic flame temperature, or flame speed observed using the true chemical kinetics~\citep{Bhattacharjee2003, Peters2008, Pepiot2008, Kelly2021}.
These reduced mechanisms are typically formulated using a combination of theory and intuition, leaving unresolved chemistry, resulting in uncertainties in the relevant rate parameters~\cite{Hakim2016}.
Selecting a functional form of the modeled reaction rate terms that lead to reliable parameter estimation is highly desirable ~\citep{Vajda1989, Deussen2022}.
This means that for high confidence in parameter estimation and thereby model prediction, the underlying parameterization of reaction rate terms must exhibit high practical identifiability.

Shock tube ignition is a canonical experiment used to develop and validate combustion reaction mechanism~\cite{Davidson2004}.
In such experiments, the reactant mixture behind the reflected shock experiences elevated temperature and pressure, followed by mixture combustion.
An important quantity of interest in such experiments is the time difference between the onset of the reflected shock and the ignition of the reactant mixture, defined as the ignition delay $t_{ign}$~\cite{Chaos2010}.
Ignition delay is characterized as the time of maximum heat release or steepest change in reactant temperature and is therefore a key physio-chemical property for combustion systems.
\begin{figure}[ht]
    \centering
    \includegraphics[width = 0.6\textwidth]{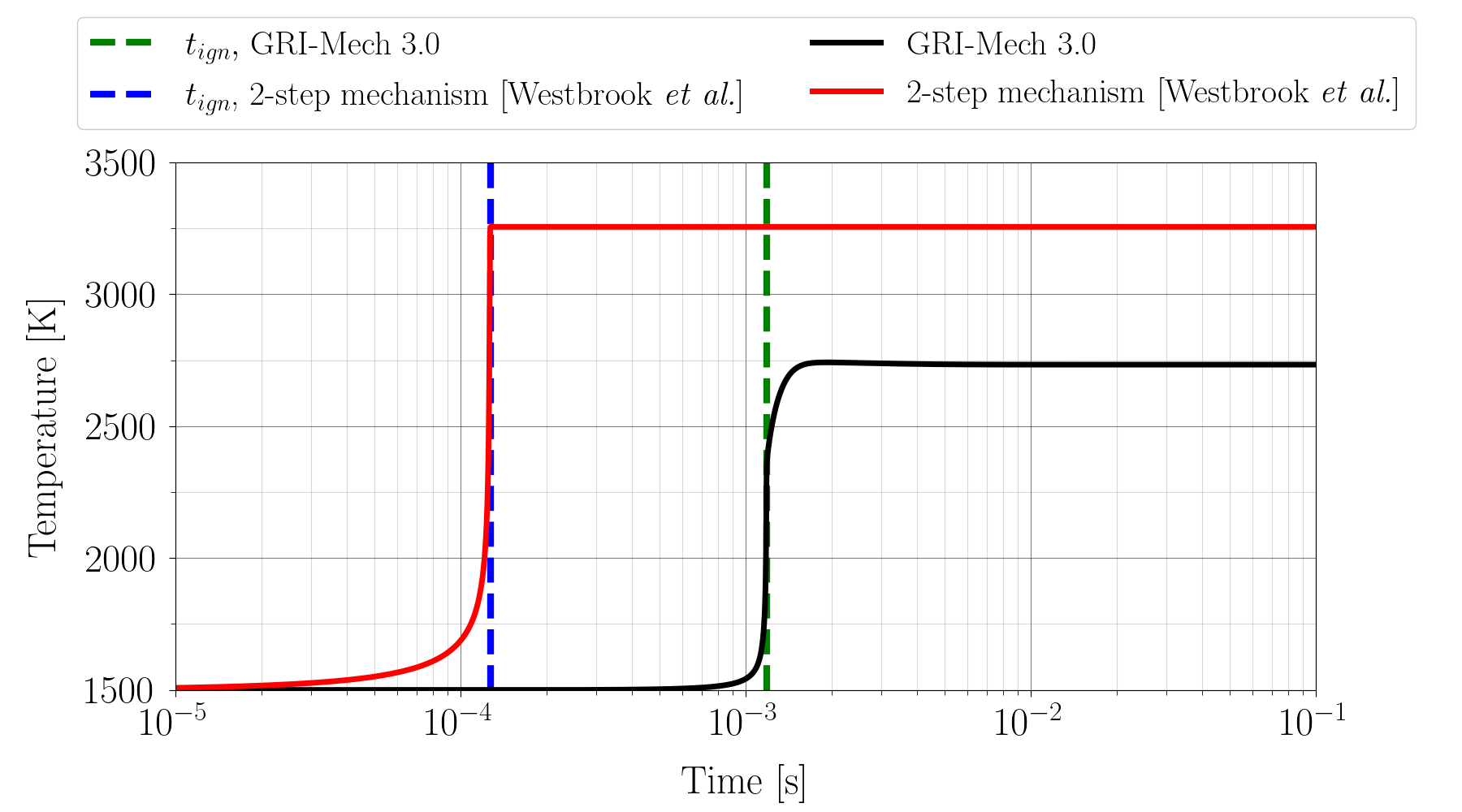}
    \caption{Temperature evolution for stoichiometric methane-air combustion at $T_o$ = $\SI{1500}{\kelvin}$, $P_o$ = $\SI{100}{\kilo\pascal}$ by means of 2-step mechanism~\cite{Westbrook1981} in comparison with GRI-Mech 3.0.
    Ignition delay time $t_{ign}$ at which mixture releases maximum heat is under-predicted by nearly an order of magnitude by the 2-step mechanism.}
    \label{fig:combustion_temperature_prediction}
\end{figure}
To illustrate the practical identifiability framework we will consider stoichiometric methane-air combustion in a shock tube under an adiabatic, ideal-gas constant pressure ignition assumption.
Typically, the chemical kinetics capturing detailed chemistry of methane-air ignition is computationally expensive due to hundreds of associated reactions.
To model the reaction chemistry, consider the classical 2-step mechanism proposed by~\citet{Westbrook1981} that accounts for the incomplete oxidation of methane.
This reduced mechanism consists of a total of 6 species (5 reacting and 1 inert species, namely, $\mathrm{N_2}$) and 2 reactions (1 reversible), thus drastically reducing the cost of evaluating the chemical kinetics.
The reactions involved in this reduced chemical kinetics model are
\begin{eqnarray}
    \ce{CH_4} + \frac{3}{2} \ce{O_2} \xrightarrow[]{k_{1}} \ce{CO} + 2 \ce{H}_{2}\ce{O},  \\
    \ce{CO} + \frac{1}{2} \ce{O}_2 \xrightleftharpoons[k_{2b}]{ k_{2f} } \ce{CO}_2,
\end{eqnarray}
where the overall reaction rates are temperature-dependent and are modeled using the Arrhenius rate equation as
\begin{eqnarray}
    k_1 &\triangleq& Ae^{ \frac{-48400}{RT} }[\ce{CH_4}]^{-0.3}[\ce{O}_2]^{1.3},\\
    k_{2f} &\triangleq& 3.98\times 10^{14}e^{ \frac{-40000}{RT} }[\ce{CO}][\ce{H_2O}]^{0.5}[\ce{O}_2]^{0.25},\\
    k_{2b} &\triangleq& 5\times 10^{8}e^{ \frac{-40000}{RT} }[\ce{CO_2}],
\end{eqnarray}
where $A=2.8\times 10^9$ is the pre-exponential factor, $R$ is the ideal gas constant, and $T$ is the temperature in Kelvin.
To solve the resulting reaction equations \texttt{CANTERA v2.6.0}~\cite{Cantera} is used.
Figure~\ref{fig:combustion_temperature_prediction} illustrates the temperature evolution using the 2-step mechanism and GRI-Mech 3.0~\cite{Smith1999}
for an initial temperature $T_o$ = $\SI{1500}{\kelvin}$, initial pressure $P_o$ = $\SI{100}{\kilo\pascal}$, and at a stoichiometric ratio $\phi=1$.
The GRI-Mech 3.0 mechanism consists of detailed chemical kinetics with 53 species and 325 reactions.
As noticed, the 2-step mechanism under-predicts the ignition delay by nearly an order of magnitude.
To improve the predictive capabilities of the 2-step mechanism a functional dependency for the pre-exponential factor can be introduced as $\log A = \fancy{G} (T_o, \phi)$,
where

\begin{equation}
    \fancy{G} (T_o, \phi) \triangleq 18 + \theta_1 + \tanh{(\theta_2 + \theta_3*\phi)\frac{T_o}{1000}}.
   \label{eqn:pre_exp_pamareterization}
\end{equation}
Here, $\theta_1, \theta_2$, and $\theta_3$ are the uncertain model parameters.
Similar parameterization was used by~\citet{Hakim2016} for $n$-dodecane reduced chemical kinetics.
It should be noted that while a more expressive functional form for the pre-exponential factor can be chosen in (\ref{eqn:pre_exp_pamareterization}), the goal of the framework is to ascertain practical identifiability.
For parameter estimation, consider the detailed GRI-Mech 3.0 to be the `exact solution' to the combustion problem which can then be used to generate the data.
Consider logarithm of ignition temperature at $T_{o} = \SI{1100}{}, \SI{1400}{}, \SI{1700}{}$ and $\SI{2000}{\kelvin}$ at $\phi=1.0$ and $P_o$ = $\SI{100}{\kilo\pascal}$ as the available data for model calibration.
Assume an uncorrelated measurement noise $\Gamma=\sigma_{\xi}^2\mathbb{I}$ with $\sigma_{\xi}^2 = 0.1$.

\subsubsection{Parameter Identifiabiliy}
The practical identifiability framework is now applied to the methane-air combustion problem to examine the identifiability of the model parameters in (\ref{eqn:pre_exp_pamareterization}) before any controlled experiments are conducted.
Consider an uncorrelated prior distribution for the model parameters as $\theta_1\sim\mathcal{N} (0, 1); \theta_2\sim\mathcal{N} (0, 1); \theta_3\sim\mathcal{N} (0, 1)$.
Such priors result in pre-exponential factors in the order similar to those reported by~\citet{Westbrook1981}, and are therefore considered suitable for the study.
Similar to~\S\ref{sec:linear_gaussian} historical estimates of the model parameters are not considered for examining identifiability.
\begin{figure}[ht]
    \centering
    \includegraphics[scale=0.28]{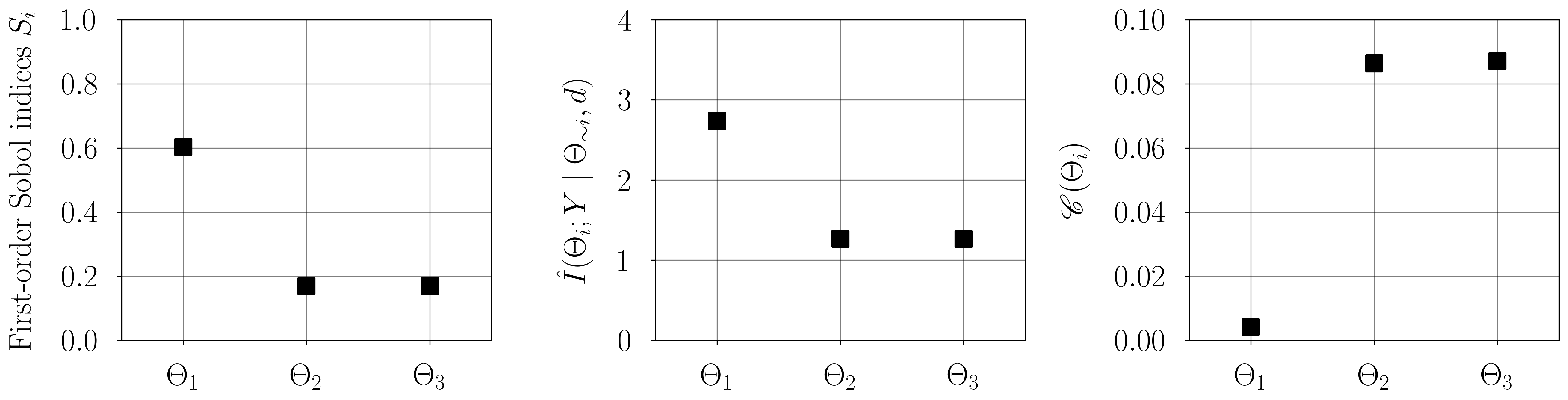}
    \caption{First-order Sobol effect indices (left), information gain (center), and information gain equivalent variance $\fancy{C}(\Theta_i)$ (right) for methane-air combustion model.
    Sobol indices show that the largest variability in the output of the statistical model is to uncertainty in $\Theta_1$; $\Theta_2$ and $\Theta_3$ exhibit similar variabilities.
    Variable $\Theta_{1}$ exhibits the information gain and therefore highest practical identifiability; $\Theta_2$ and $\Theta_3$ have similar information gain.
    Variable $\Theta_1$ exhibits the lowest measurement uncertainty for the direct observation model, followed by similar uncertainty for $\Theta_2$ and $\Theta_3$.
    }
    \label{fig:combustion/iden_vs_sobol}
\end{figure}

The first-order Sobol indices, estimated information gain, and information gain equivalent variance $\fancy{C}(\Theta_i)$ are shown in figure~\ref{fig:combustion/iden_vs_sobol}.
The information gain is estimated using $n_{\text{outer}}=12000$ Monte-Carlo samples, and $n_{\text{inner}}=5$ quadrature points.
Examining the first-order Sobol indices (\suppinfo[figure~\ref{fig:combustion_sobol_convergence}] for convergence study), the output of the forward model exhibits the largest variability due to uncertainty in the variable $\Theta_1$.
Followed by similar variability in the model output with respect to $\Theta_2$ and $\Theta_3$.
The largest information gain is observed for the variable $\Theta_1$, followed by similar gains for $\Theta_2$ and $\Theta_3$.
This means that $\Theta_1$ will have the highest practical identifiability, followed by a much lower identifiability for $\Theta_2$ and $\Theta_3$.
Using the hypothetical direct observation model as described in~\S\ref{sss:interpreting_identifiability}, the variable $\Theta_1$ with the largest practical identifiability exhibits the lowest measurement uncertainty, followed by similar uncertainty for $\Theta_2$ and $\Theta_3$.
\begin{figure}[ht]
    \centering
    \includegraphics[scale=0.05]{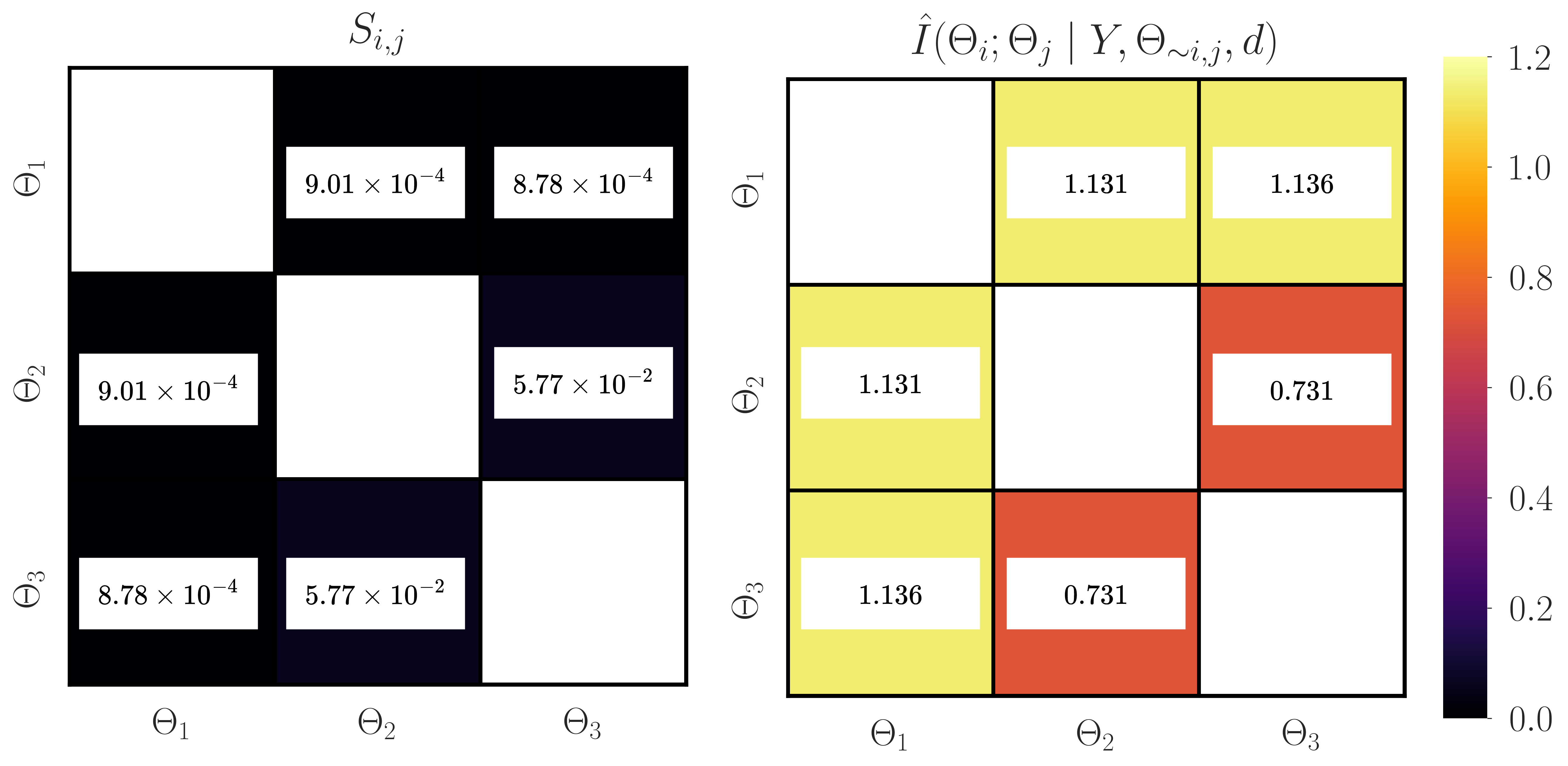}
    \caption{Second-order Sobol indices (left) and estimated parameter dependencies (right) for methane-air combustion model.
    The trend $S_{2, 3} > S_{1, 2} \approx S_{1, 3}$ suggests that there are underlying interactions between the parameters $\Theta_2$ and $\Theta_3$.
    Pairs ($\Theta_1$, $\Theta_2$) and ($\Theta_1$, $\Theta_3$) have nearly the same dependencies on one another. Pair ($\Theta_2$, $\Theta_3$) exhibit low dependencies.
    }
    \label{fig:combustion/estimated_dependencies}
\end{figure}

Figure~\ref{fig:combustion/estimated_dependencies} shows the second-order Sobol indices and estimated parameter dependencies.
The second-order Sobol indices (\suppinfo[figure~\ref{fig:combustion_sobol_convergence}] for convergence study) follow the trend $S_{2, 3} > S_{1, 2} \approx S_{1, 3}$, suggesting that there are underlying interactions between the parameters $\Theta_2$ and $\Theta_3$.
As observed, the low identifiability of $\Theta_2$ and $\Theta_3$ suggested in figure~\ref{fig:combustion/iden_vs_sobol} is primarily due to the underlying dependencies between pairs $(\Theta_1, \Theta_2)$ and $(\Theta_1, \Theta_3)$.
To estimate the parameter dependencies $n_{\text{outer}}=12000$ Monte-Carlo samples, and $n_{\text{inner}}=5$ and $10$ quadrature points are used for single and two-dimensional integration space, respectively.
Similar magnitude of parameter dependencies obtained for the pairs $(\Theta_1, \Theta_2)$ and $(\Theta_1, \Theta_3)$ in addition to similar information gain for $\Theta_2$ and $\Theta_3$ also suggest underlying symmetry with respect to $\Theta_1$.
This means that the interchange of $\Theta_2$ and $\Theta_3$ will not affect the output of the statistical model, which can be clearly seen in (\ref{eqn:pre_exp_pamareterization}) for $\phi=1$.
This is also evident from the second-order Sobol indices which suggest that there is a combined effect on the output of the statistical model due to interactions between $\Theta_2$ and $\Theta_3$.

\begin{figure}[ht]
    \centering
    \includegraphics[scale=0.078]{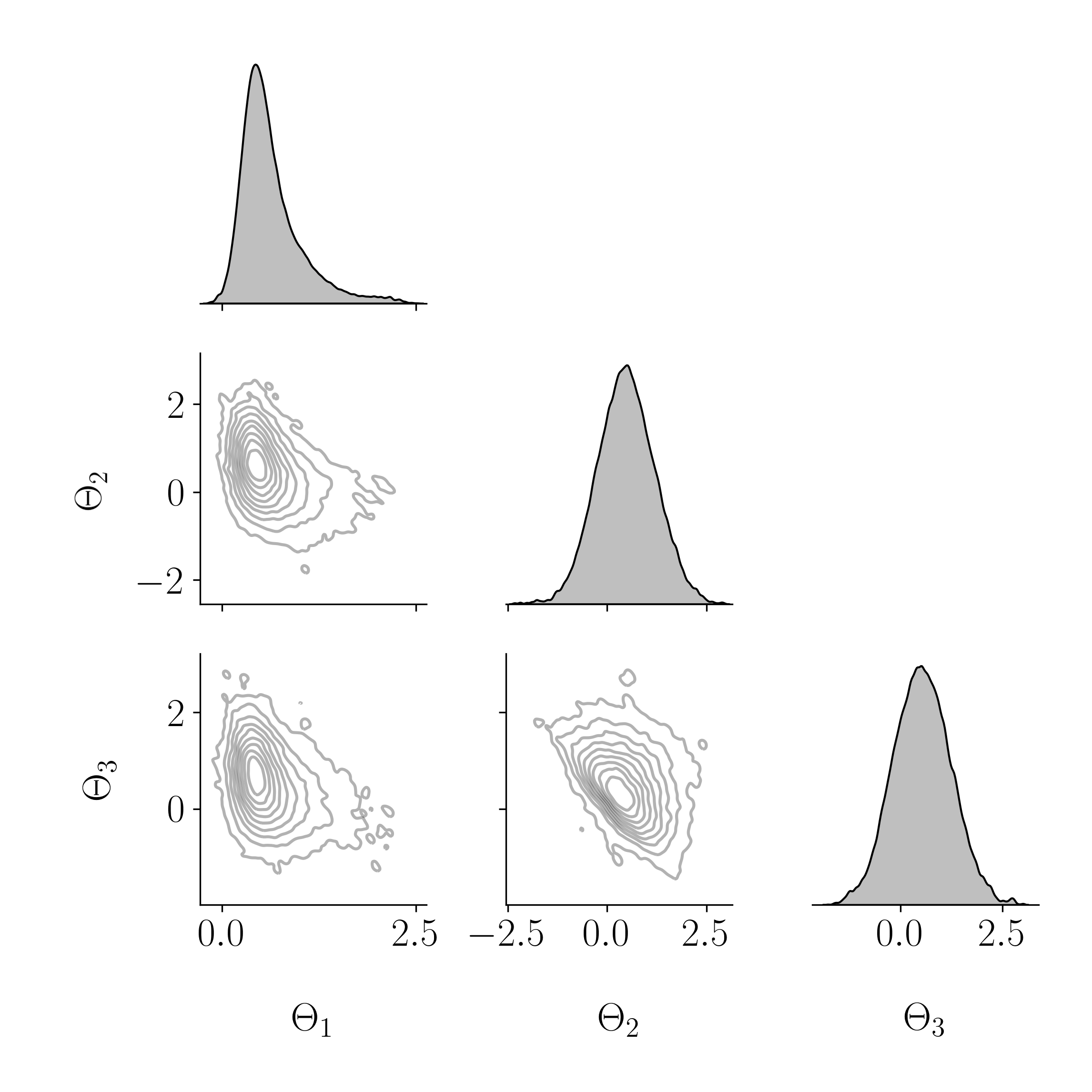}
    \includegraphics[scale=0.25]{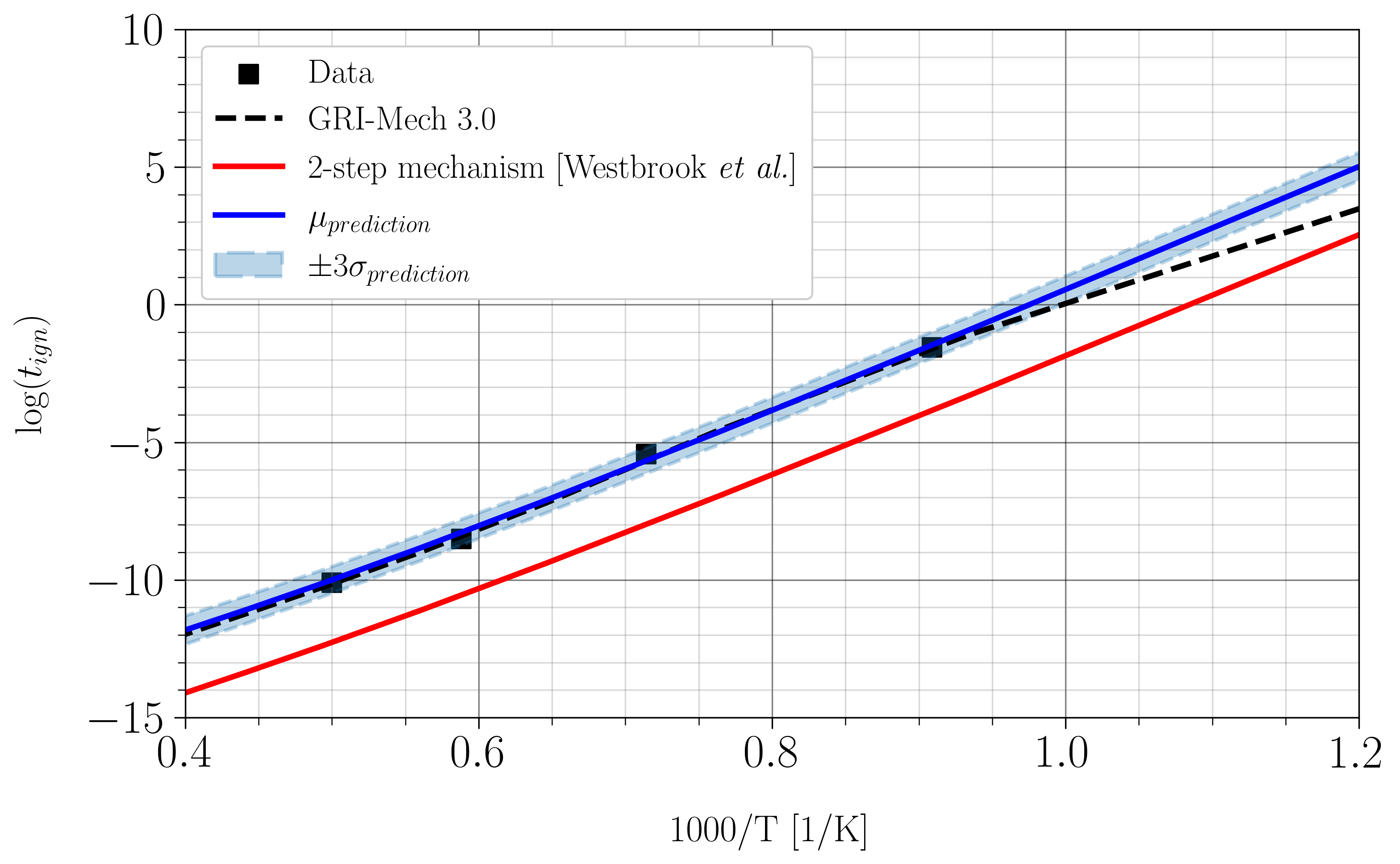}
    \caption{Correlation plot for samples obtained from the posterior distribution (left) and the obtained aggregate posterior prediction (right) for the methane-air combustion model.
    Correlation plots do not reveal any relations among variables.
    Aggregate posterior prediction agrees well with the data and exhibits high certainty.
    }
    \label{fig:combustion/posterior_prediciton}
\end{figure}
\subsubsection{Parameter Estimation}
Now, let us consider the parameter estimation problem which seeks $p(\theta\mid y)$, that is the posterior distribution.
Typically, a closed-form expression for the posterior distribution is not available due to the non-linearities in the forward model or the chosen family of the prior distribution.
As an alternative, sampling-based methods such as Markov Chain Monte Carlo (MCMC) that seek samples from an unnormalized posterior have gained significant attention.
These methods construct Markov chains for which the stationary distribution is the posterior distribution.
The Metropolis-Hastings algorithm is an MCMC method that can be used to generate a sequence of samples from any given probability distribution~\cite{Tierney1994}.
The adaptive Metropolis algorithm is a powerful modification to the Metropolis-Hastings algorithm and is used here to sample from the posterior distribution~\cite{Haario2001}.

Figure~\ref{fig:combustion/posterior_prediciton} illustrates the correlation between samples obtained using the Adaptive Metropolis algorithm and the obtained aggregate posterior prediction for ignition delay time.
Any correlation (linear) is not observed between the variables; however, the joint distribution between pairs $(\Theta_1, \Theta_2)$ and $(\Theta_1, \Theta_3)$ show similarities.
These similarities were also observed during the a priori analysis quantifying parameter dependencies as shown in figure~\ref{fig:combustion/estimated_dependencies}.
\begin{figure}[ht]
    \centering
    \includegraphics[scale=0.195]{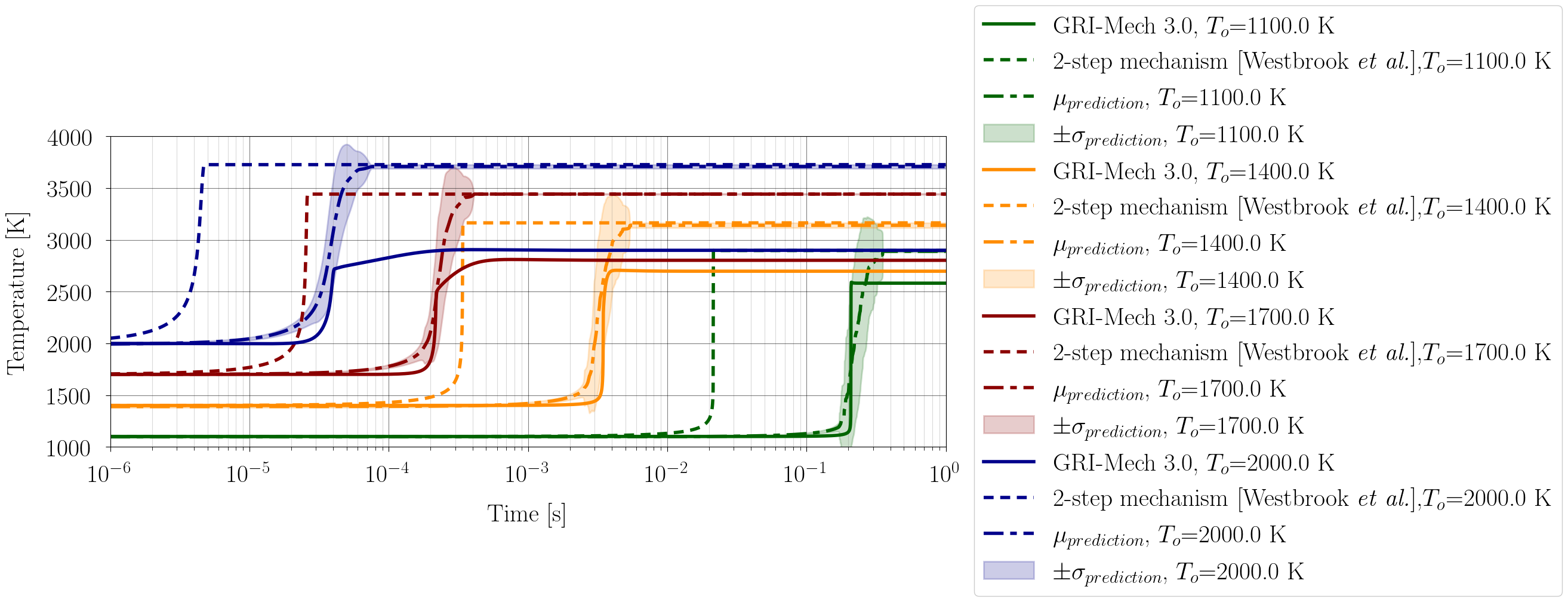}
    \caption{Aggregate temperature evolution for methane-air combustion model.
    Aggregate prediction agrees well with the GRI-Mech 3.0 detailed mechanism.}
    \label{fig:combustion/temperature_evolution}
\end{figure}
\begin{figure}[h]
    \centering
    \includegraphics[scale=0.185]{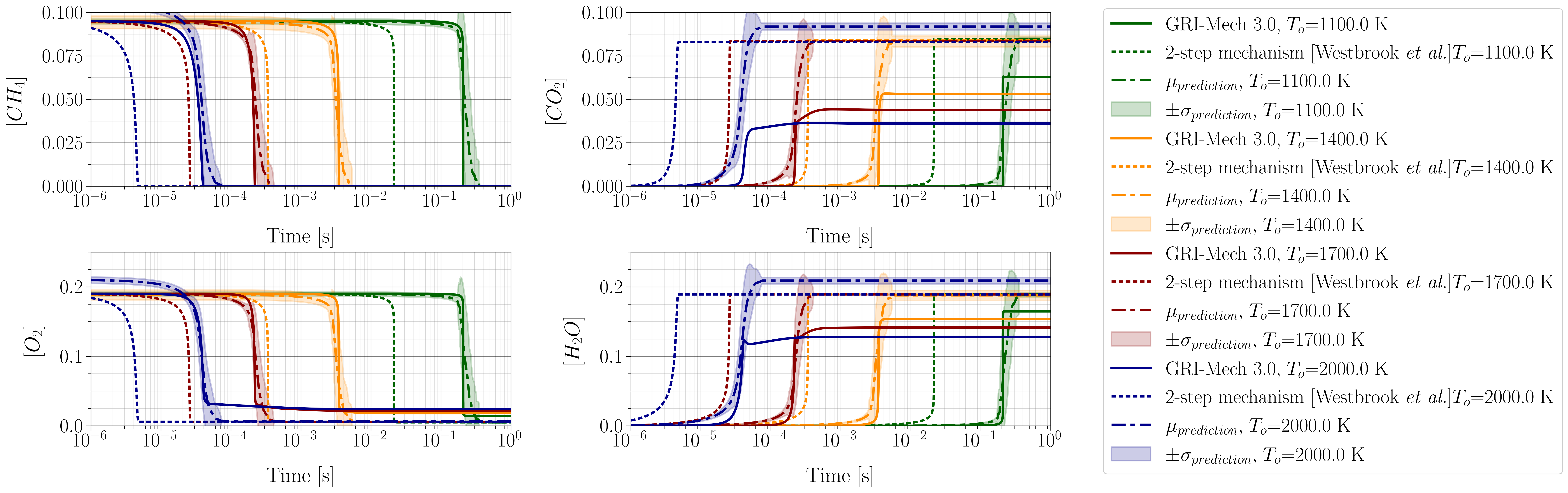}
    \caption{Aggregate species concentration evolution for methane-air combustion model.
    Aggregate prediction agrees well with the GRI-Mech 3.0 detailed mechanism.}
    \label{fig:combustion/species_concentration}
\end{figure}
The obtained aggregate prediction shows dramatic improvement over the 2-step mechanism in predicting ignition delay time over a wide range of temperatures.
Using a functional form as (\ref{eqn:pre_exp_pamareterization}) for the pre-exponential factor also improved the mixture temperature evolution, as shown in figure~\ref{fig:combustion/temperature_evolution}.
However, the adiabatic flame temperature, which is defined as the mixture temperature upon reaching equilibrium, is still being over-predicted.
An improvement in the prediction of the evolution of species concentration over time is also noticed, as shown in figure~\ref{fig:combustion/species_concentration}.



\section{Concluding remarks and perspectives}
\label{sec:conclusion}
Examining the practical identifiability of statistical models is useful in many applications, such as parameter estimation, model-form development, and model selection.
Estimating practical identifiability prior to conducting controlled experiments or parameter estimation studies can assist in a choice of parametrization that can be associated with a high degree posterior certainty, thus improving confidence in estimation and model prediction.

In this work, a novel information-theoretic approach based on conditional mutual information is presented to assess global practical identifiability of a statistical model in a Bayesian framework.
The proposed framework examines the expected information gain for each parameter from the data before performing controlled experiments.
Parameters with higher information gain are characterized by having higher posterior certainty, and thereby have higher practical identifiability.
The adopted viewpoint is that the practical identifiability of a parameter does not have a binary answer, rather it is the relative practical identifiability among parameters that is useful in practice.
In contrast to previous numerical approaches used to study practical identifiability, the proposed approach has the following notable advantages: first, no controlled experiment or data is required to conduct the practical identifiability analysis; second, different forms of uncertainties, such as model-form, parameter, or measurement can be taken into account; third, the framework does not make assumptions about the distribution of the data and parameters as in the previous methods; fourth, the estimator provides knowledge about global identifiability and is therefore not dependent on a particular realization of the parameters.
To provide a physical interpretation to practical identifiability in the context of examining information gain for each parameter, an information gain equivalent variance for a direct observation model is also presented.
The practical identifiability framework is then extended to examine dependencies among parameter pairs.
Even if an individual parameter exhibits poor practical identifiability characteristics, it can belong to an identifiable subset such that parameters within the subset have functional relationships with one another.
Parameters within such an identifiable subset have a combined effect on the statistical model and can be collectively identified.
To find such subsets, a novel a priori estimator is proposed to quantify the expected dependencies between parameter pairs that emerge a posteriori.

To illustrate the framework, two statistical models are considered: (a) a linear Gaussian model and (b) a non-linear methane-air reduced kinetics model.
For the linear Gaussian model, it is shown that parameters with large information gain and low parameter dependencies can be estimated with high confidence.
The variance-based global sensitivity analysis (GSA) also illustrates that parameter sensitivity is necessary for identifiability.
However, as conclusively shown, the inability of variance-based GSA to capture different forms of uncertainties can lead to unreliable estimates for practical identifiability.
The information gain equivalent variance obtained using a direct observation model shows that parameters with high practical identifiability will be associated with low measurement uncertainty if observed directly.
In the case of the methane-air reduced kinetics model, it is shown that parameters with large dependencies can have low information gain and therefore low practical identifiability.
Further, the proposed estimator can capture non-linear dependencies and reveal structures within the parameter space before performing controlled experiments.
Such non-linear dependencies cannot be observed when considering a posteriori parameter correlations, as only linear relations can be well understood.

\section*{Declarations}

\begin{itemize}
\item Funding: The authors disclose support for the research of this work from OUSD(RE) Grant No: N00014-21-1-295.
\item The authors declare no competing interests.
\item Ethics approval
\item Consent to participate
\item Consent for publication
\item Availability of data and materials: All data generated or analyzed during this study are included in this published article.
\item Code availability: Code related to the experiments performed in this article is available at \url{https://github.com/sahilbhola14/GIM}
\item SB primarily developed the methodology and generated numerical results. KD formulated the  problem statement and provided guidance for the work.
\end{itemize}

\nocite{Herman2017, Iwanaga2022}
\bibliography{sn-bibliography}

\begin{appendices}

    \section{Global Sobol Sensitivity Analysis}\label{sec:sobol_index}
Variance-based sensitivity analysis has been widely used to determine the most influential parameters by examining variability in the forward model response as a result of uncertainty in the model parameters~\cite{Saltelli2000}.
This is primarily based on the idea that parameters which cause large variability in model output are most influential and therefore relevant in the context of parameter estimation.
Note that while sensitivity analysis comments on the degree of variability in the model output due to parameter uncertainty, it does not quantify the degree of confidence in parameter estimation and therefore practical identifiability.
While parameter sensitivity is necessary for identifiability it is not sufficient~\cite{Wu2019}.

Global sensitivity analysis based on Sobol indices apportions the variability in the response of the forward model only with respect to uncertainty in the parameters.
Consider the statistical model in (\ref{eqn:statistical_model}) with no measurement uncertainty, that is $y=\fancy{F}(\theta, d)$.
Without the loss of generality, consider the model output as a scalar.
For such a model the total output variability can be uniquely decomposed as

\begin{equation}
\var{\Theta}{Y}\triangleq\sum_{i=1}^m V_i+\sum_{i<j}^m V_{i j}+\cdots+V_{12 \ldots m},
\end{equation}
under the assumption that all $\Theta_{i}\in \Theta$ are mutually statistically independent~\cite{Saltelli2008}.
Here

\begin{align}
    V_i & \triangleq\var{\Theta_i}{\expect{\Theta_{\sim i}}{Y\mid \Theta_i}},\label{eqn:sobol_numerator}\\
    V_{i j} & \triangleq\var{\Theta_i, \Theta_j}{\expect{\Theta_{\sim i, j}}{Y\mid \Theta_i, \Theta_j}}-V_i-V_j,
\end{align}
\noindent
and so forth.
Variance-based Sobol indices can then be defined as

\begin{align}
    S_i &\triangleq \frac{V_i}{\var{\Theta}{Y}},\label{eqn:main_effect_sobol}\\
    S_{ij} &\triangleq \frac{V_{ij}}{\var{\Theta}{Y}},
\end{align}
and so forth.
As a result

\begin{equation}
\sum_{i=1}^m S_i+\sum_{i<j}^m S_{i j}+\cdots+S_{12 \ldots m}=1.
\end{equation}
The $S_{i}$ terms are called the first-order or the main effect Sobol indices, which apportions the output variability due to uncertainty in $\Theta_i$.
The multi-index terms such as $S_{ij}$ are the higher order interaction terms that apportion the output variability due to the combined uncertainty in parameter sets.
For example, the second-order Sobol indices $S_{ij}$ apportions the output variability due to the interaction between parameter pairs.
Note, in the case of vector model output the expected Sobol indices can be evaluated. In most practical applications, the first-order Sobol indices in (\ref{eqn:main_effect_sobol}) are sufficient to ascertain parameter relevance from the perspective of model sensitivity.
In the present work, the Sobol indices are estimated using \texttt{SALib}~\citep{Herman2017, Iwanaga2022}.
It requires $N*(2m + 2)$ model evaluations, where $N$ is the number of parameter samples generated and $m$ is the number of parameters in the statistical model.


    \section{Linear Gaussian Model}\label{sec:linear_gaussian_app}
For a random variable $Z$ distributed normally as $\mathcal{N}(\mu, \Sigma)$ the differential entropy has a closed form expression

\begin{equation}
H(Z)=\frac{n}{2} \ln (2 \pi)+\frac{1}{2} \ln |\Sigma|+\frac{1}{2} n,
\label{eqn:differential_entropy_gaussian}
\end{equation}
where $\mu\in\real{n}$ and $\Sigma\in\real{n\times n}$~\cite{Cover1999}.

Consider the differential entropy formulation for information gain in (\ref{eqn:information_content_differential_form}).
To evaluate information gain for a linear Gaussian model in (\ref{eqn:linear_gaussian_model}), closed-form expressions are available for the densities $p(\theta_i, \theta_{\sim i}\mid d)$, $p(\theta_{\sim i}, y\mid d)$, $p(\theta_{\sim i}\mid d)$, and $p(\theta_i, \theta_{\sim i\mid d}, y\mid d)$ such that (\ref{eqn:differential_entropy_gaussian}) can be used to find the true information gain.

Similarly, consider the differential form for evaluating the parameter dependencies in (\ref{eqn:parameter_dependence_differential_form}).
Closed form expressions are available for the densities $p(\theta_i, \theta_{\sim i, j}, y\mid d)$, $p(\theta_j, \theta_{\sim i, j}, y\mid d)$, $p(\theta_{\sim i, j}, y\mid d)$, and $p(\theta_i, \theta_j \theta_{\sim i, j}\mid d)$ such that true parameter dependencies can be evaluated using (\ref{eqn:differential_entropy_gaussian}).

    \section{Convergence of Sobol Indices}\label{sec:sobol_convergence}
\subsection{Linear Gaussian Model}
\begin{figure}[h!]
    \centering
    \includegraphics[width = 0.9\textwidth]{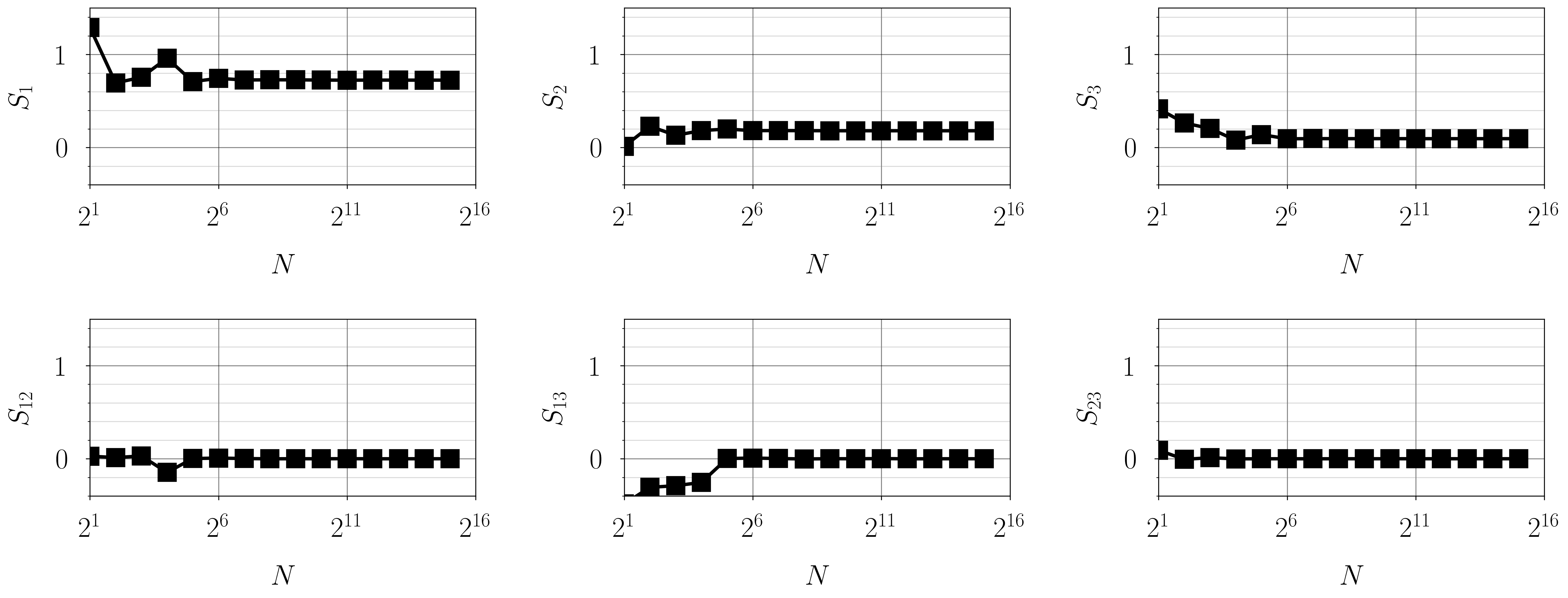}
    \caption{Convergence of first-order Sobol indices (top) and second-order Sobol indices (bottom) with number of parameter samples $N$ (\suppinfo[\S\ref{sec:sobol_index}]) for linear Gaussian model.}
    \label{fig:linear_gaussian_sobol_convergence}
\end{figure}

\subsection{Methane Chemical Kinetics Model}
\begin{figure}[h!]
    \centering
    \includegraphics[width = 0.9\textwidth]{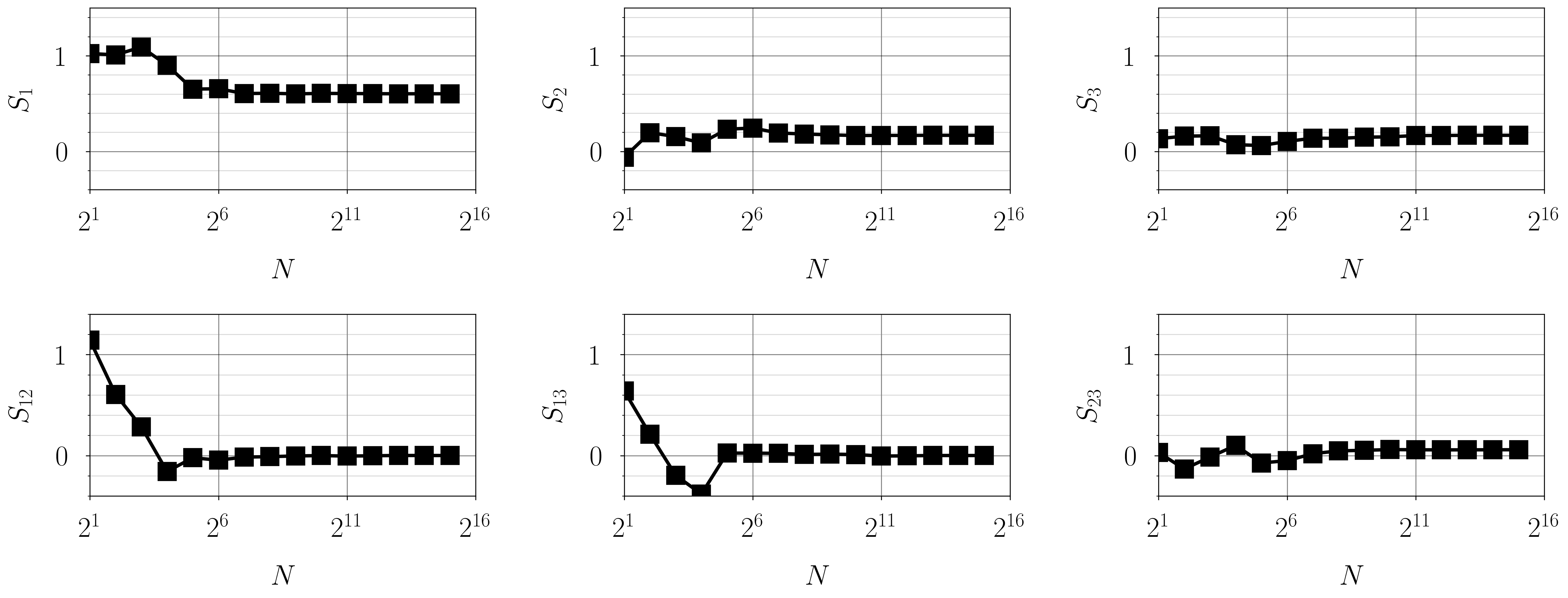}
    \caption{Convergence of first-order Sobol indices (top) and second-order Sobol indices (bottom) with number of parameter samples $N$ (\suppinfo[\S\ref{sec:sobol_index}]) for methane-air combustion model.}
    \label{fig:combustion_sobol_convergence}
\end{figure}





\end{appendices}

\end{document}